\documentclass[sigconf]{acmart}

\usepackage{soul}
\usepackage{xspace}     
\usepackage{booktabs}    
\usepackage{array}       
\usepackage{xcolor}      
\usepackage{colortbl}    
\usepackage{graphicx}    
\usepackage{menukeys}    
\usepackage{dirtytalk}   
\usepackage{tabularx}    
\usepackage{enumitem}    

\usepackage{etoolbox}
\usepackage{quoting}

\newenvironment{myblockquote}{%
  \vspace{0.5em}
  \small
  \begin{quoting}[leftmargin=0.5em, rightmargin=0.5em, indentfirst=false]%
  \setlength{\parindent}{0pt}%
  \itshape
}{%
  \end{quoting}%
  \vspace{0.5em}
}

\newcommand{\sysname}{Street\-Reader\-AI\xspace}
\newcommand{\myquote}[1]{{\textit{\say{#1}}}}    

\ifdefined\keys
\else
    \newcommand{\keys}[1]{\texttt{#1}}
\fi

\newif\ifshowrevisions
\showrevisionsfalse 

\newcommand{\revised}[1]{%
    \ifshowrevisions%
        \textcolor{red}{#1}%
    \else%
        #1%
    \fi%
}

\AtBeginDocument{%
  }

\copyrightyear{2025}
\acmYear{2025}
\setcopyright{cc}
\setcctype{by}
\acmConference[UIST '25]{The 38th Annual ACM Symposium on User Interface Software and Technology}{September 28-October 1, 2025}{Busan, Republic of Korea}
\acmBooktitle{The 38th Annual ACM Symposium on User Interface Software and Technology (UIST '25), September 28-October 1, 2025, Busan, Republic of Korea}\acmDOI{10.1145/3746059.3747756}
\acmISBN{979-8-4007-2037-6/2025/09}

\begin{document}

\title[\sysname]{\sysname: Making Street View Accessible Using Context-Aware Multimodal AI}


\author{Jon E. Froehlich}
\affiliation{%
  \institution{Google Research}
  \country{} 
  }
\email{jfroehlich@google.com}

\author{Alex Fiannaca}
\affiliation{%
  \institution{Google DeepMind}
  \country{}
  }
\email{afiannaca@google.com}

\author{Nimer Jaber}
\affiliation{%
  \institution{Google}
  \country{}
  }
\email{nimer@google.com}

\author{Victor Tsaran}
\affiliation{%
  \institution{Google}
  \country{}
  }
\email{vtsaran@google.com}

\author{Shaun Kane}
\affiliation{%
  \institution{Google Research}
  \country{}
  }
\email{shaunkane@google.com}

\renewcommand{\shortauthors}{Froehlich et al.}

\begin{abstract}
Interactive streetscape mapping tools such as Google Street View (GSV) and Meta Mapillary enable users to virtually navigate and experience real-world environments via immersive 360° imagery but remain fundamentally inaccessible to blind users. We introduce \textit{\sysname}, the first-ever accessible street view tool, which combines context-aware, multimodal AI, accessible navigation controls, and conversational speech. With \sysname, blind users can virtually examine destinations, engage in open-world exploration, and virtually tour any of the over 220 billion GSV images across 100+ countries. We iteratively designed \sysname with a mixed-visual ability team and performed an evaluation with eleven blind users. Our findings demonstrate the value of an accessible street view in supporting POI investigations and remote route planning. We close by enumerating key guidelines for future work.




\end{abstract}



\begin{teaserfigure}
  \includegraphics[width=\textwidth]{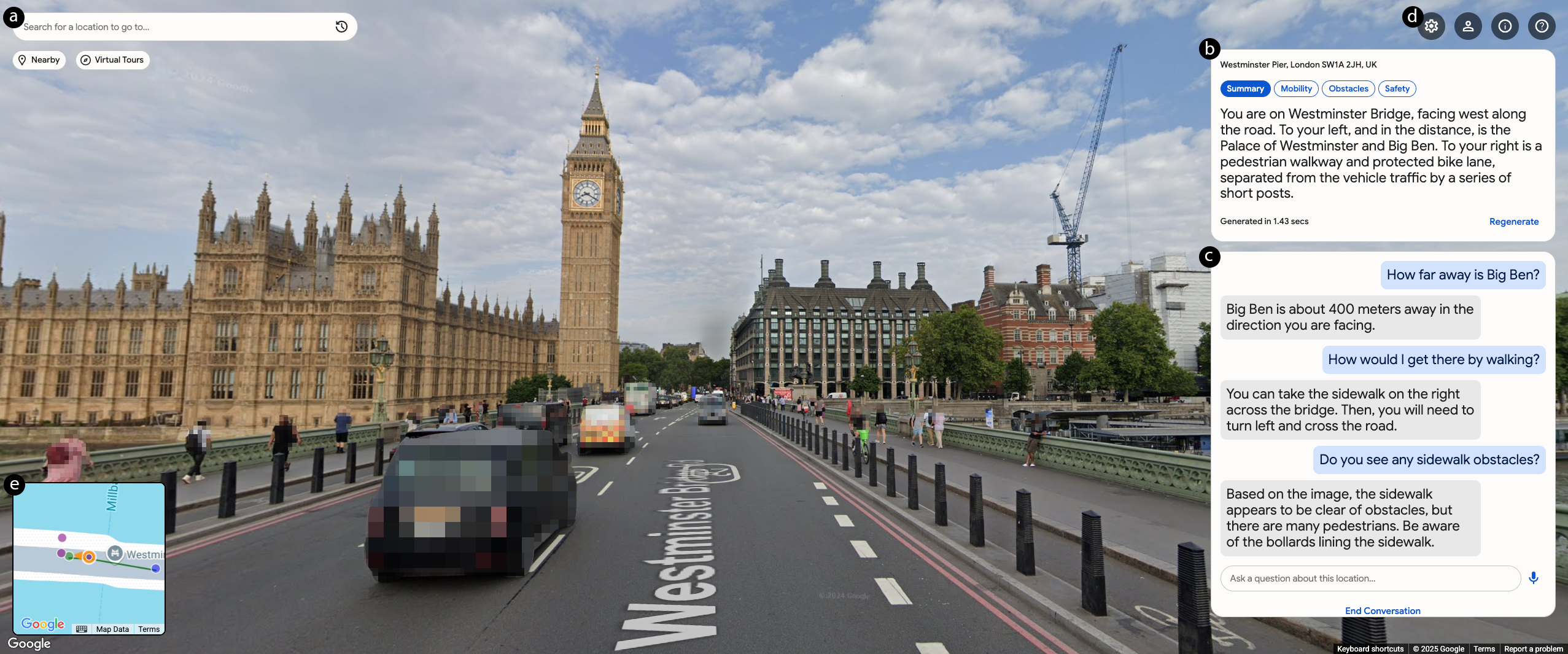}
  \vspace{-2.0em} 
  \caption{We introduce \sysname, an accessible streetscape mapping prototype that uses context-aware AI and accessible navigation controls for blind and low-vision users. (a) Users can search for and select locations, (b) trigger AI-based descriptions, or (c) chat with a multimodal AI agent about the scene and local geography while virtually exploring the world.\textsuperscript{{\scriptsize 1}}}
  \Description{Shows a screenshot of \sysname in England with Big Ben in the distance. AI-generated description reads: 'You are on Westminster Bridge, facing west along the road. In front of you, the bridge has cars traveling in both directions and a protected sidewalk with pedestrians on both sides. In the distance, you can see the Houses of the Parliament and the clock tower known as Big Ben.' The AI chat then shows the user asking about the distance and walking directions.}
  \label{fig:teaser}
\end{teaserfigure}



\maketitle

\begin{figure*}[t!] 
  \centering 
  \includegraphics[width=\textwidth]{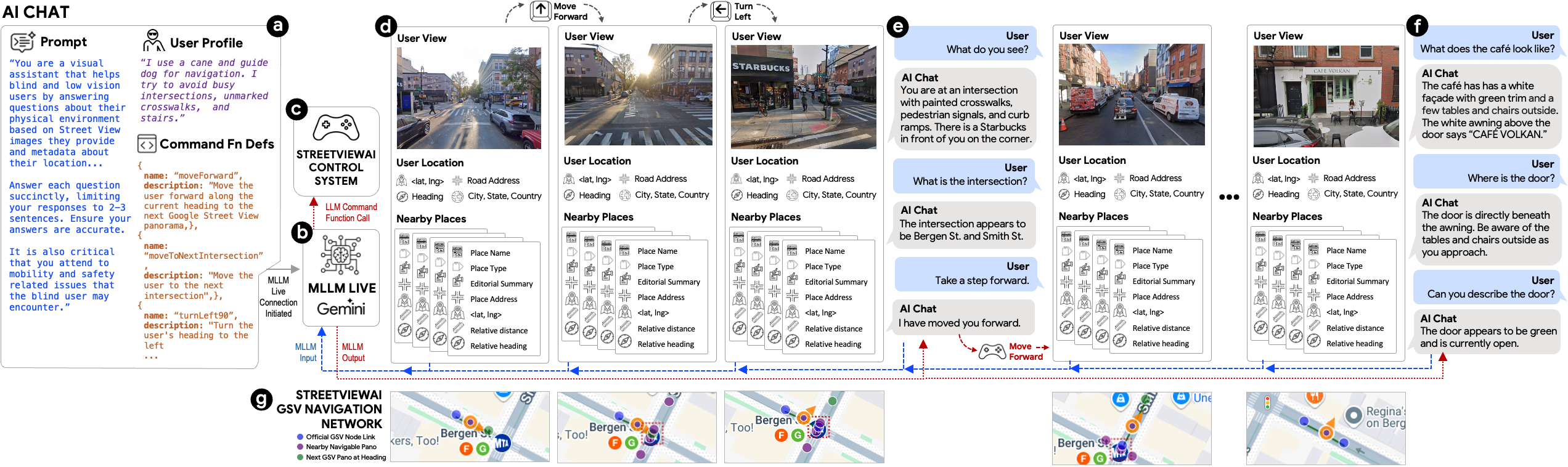}
  \vspace{-2.0em} 
  \caption{\revised{\sysname's context-aware multimodal \textit{AI Chat}. (a) The text prompt, optional user profile, and command function definitions are sent on "MLLM live chat" initiation. (b) The MLLM Live module and (c) panning/movement control system. (d) Once a chat is initiated, \sysname uploads the user's current view and geographic context (\textit{e.g.,} <lat,lng>, current heading, nearby places)---seamlessly updated every view change or movement. (e, f) Example context-aware AI chats. (g) The custom GSV-based navigational network. In this example, the user is virtually exploring a walk from a transit station to a cafe in NYC.}}
  \Description{A system diagram of AI Chat showing the multimodal input, AI chat output, and AI-based command subsystem.}
  \label{fig:ai_chat_diagram}
\end{figure*}

\section{Introduction}
\footnotetext[1]{Some streetscape figures have been slightly altered to help protect privacy.}
Interactive, digital maps have transformed how people move about the world, from route planning and turn-by-turn directions to virtual tourism. Though historically inaccessible to blind users~\cite{Froehlich_GrandChallengesInMaps_ACMInteractions2019}, recent work has significantly improved traditional two-dimensional map accessibility through audio descriptions and spatialized audio~\cite{BlindSquare, Google_GoogleMapsVoiceGuidance_2023, Microsoft_Soundscape_2023}, tactile representations~\cite{Holloway_3DPrintedMaps_ASSETS19, Holloway_TactIcons_CHI23, Nagassa_3DBuildingPlans_CHI23, Crawford_CoDesigning3DPrintedTactileMap_ASSETS24}, and dynamic haptic feedback~\cite{Ducasse_AccessInteractiveMapsForVI_BookChapter18, Paratore_ExploitingHapticAndAudio_UniversalAccess24, Zeng_InteractiveAudiohaptic_IwC14}. Despite this progress, an entire class of digital maps remains inaccessible: immersive streetscape mapping tools \cite{Google_StreetView, Meta_Mapillary, Apple_LookAround, Microsoft_Bing_Streetside}. These interactive tools enable users to virtually navigate and experience real-world environments via immersive 360° imagery---similar to a first-person video game---allowing for previously infeasible tasks, such as virtually previewing routes, visually inspecting destinations, getting a ``feel'' for a neighborhood, or even remotely visiting world-class tourist locations such as the Grand Canyon~\cite{Google_StreetView_GrandCanyon}, Petra's Treasury, or the Eiffel Tower in France.

In 2007, Google launched the first broadly available streetscape mapping tool called \textit{Street View} (GSV), which uses instrumented cars to take geo-located 360° imagery uploaded to and navigable via Google Maps~\cite{Google_StreetView}. Now, every major mapping service from Meta \textit{Mapillary}~\cite{Meta_Mapillary} to Apple \textit{Look Around}~\cite{Apple_LookAround} and Bing \textit{Streetside}~\cite{Microsoft_Bing_Streetside} provide interactive streetscape imagery. GSV alone has grown to over 220 billion images spanning 10 million miles across 100 countries \cite{Google_StreetView_Stats} while Mapillary has over 2 billion crowdsourced panoramas in 140 countries \cite{Mapillary_Stats}. Though widespread and growing, these streetscape tools remain fundamentally inaccessible to blind users due to their reliance on panoramic imagery, lack of image descriptions, and inaccessible controls.

With advances in street scene analysis~\cite{Cordts_Cityscapes_CVPR16, Chen_DeepLab_TPAMI18} and multimodal AI models~\cite{GoogleDeepMind_Gemini_2025, Liu_VisualInstructionTuning_NeurIPS23, GeminiTeam_GeminiFamily_2024}, we now have an opportunity to redefine the experience of streetscape maps for blind users. We introduce \textit{\sysname}, a new, accessible street view experience using context-aware, real-time AI and accessible navigation controls (see \autoref{fig:teaser} and video). With \sysname, users can interactively pan and move between panoramic images, learn about nearby roads, intersections, and places, hear real-time AI descriptions, and dynamically converse with a live, multimodal AI agent about the scene and local geography (\autoref{fig:ai_chat_diagram}). \sysname was designed iteratively, drawing on the experiences of two blind team members, feedback from professional designers and engineers in mixed-reality and mapping tools,  and literature in accessible first-person gaming~\cite{Atkinson_AudioQuake_Sandbox06, AFB_ShadesOfDoom, Westin_Terraformers_ICDVRAT04}, mixed-reality~\cite{Chang_WorldScribe_UIST24, GoogleDeepMind_Astra_2024}, and mapping~\cite{BlindSquare, Microsoft_Soundscape_2023}.

To evaluate \sysname and investigate the potential of accessible streetscape tools, we recruited eleven blind participants---all screen reader users---for a 90-min in-person lab study. Participants engaged in a semi-structured interview about their use of navigation tools, maps, and AI before completing a tutorial of \sysname and two sets of tasks: point-of-interest (POI) investigations and open-world navigation. Sessions completed with a brief reflection and ideas for future prototypes. From \sysname logs, observational notes, and post-task interview responses, we analyzed \textit{what} questions blind participants ask about streetscape imagery, \textit{how} well our multimodal AI models performed, \textit{how} participants navigated and oriented themselves in the virtual-physical world, and \textit{how} participants felt about our tool, key concerns, and opportunities for improvement. 

We found that users were able to effectively traverse street view images and navigate the virtual-physical world, primarily by listening to auto-generated contextual status messages (\textit{e.g.,} \myquote{Central Perk Cafe is now to your left 16 meters away}) and utilizing the multimodal AI. Interestingly, while participants could trigger AI-generated scene descriptions of their current view at any time, most preferred to interact with the context-aware \textit{AI Chat Agent} where they could ask context-based questions about the current scene, previous views, and the local geography. Of the 816 questions asked in AI Chat, 703 (86.3\%) were correctly answered, 32 were incorrect (3.9\%), and the remaining were either partially correct (26; 3.2\%) or the AI refused to answer (54; 6.6\%). Our findings also highlight challenges related to mental models of pedestrian navigation \textit{vs.} inherent limitations of streetscape imagery, a bias towards trusting AI output (even when wrong), and interaction design difficulties in creating concise audio feedback that appropriately combines complex information from navigation data, panoramic imagery, and spatial relationships.

As the first accessible street view tool, our work fundamentally advances research in accessible maps and contributes design guidelines for making immersive 360° imagery accessible, new interaction techniques to accessibly converse with a context-aware AI agent about street scenes, and enumerates new opportunities for the next-generation street view tools that use AI.
\section{Related Work}
We situate our work in research on accessible maps, mixed reality (XR), AI-generated image descriptions, and street scene analysis.

\subsection{Accessible Maps}
Making maps accessible to people who are blind or low-vision (BLV) is a grand challenge in HCI and accessibility~\cite{Froehlich_GrandChallengesInMaps_ACMInteractions2019, Brock_SIG-MakingMapsAccessible_CHI18}. To translate map data---traditionally based on visual representations---into more accessible non-visual formats, researchers have explored a variety of techniques including audio descriptions~\cite{Google_GoogleMapsVoiceGuidance_2023, Papadopoulos_ComparisonOfThreeOrientationAids_AssistiveTech17, Conway_AudioDescription_TechComm2020}, sonification (non-speech audio)~\cite{Delogu_GeoMapsDoesSonificationHelp_DisabilityAndRehab2010, Schito_ExploringMapsBySounds_IJGIS18, Heuten_Interactive3DSonificationOfCityMaps_NordiCHI06}, and tactile representations ranging from raised-line graphics \cite{Zhao_TactileFixationsRaisedLineGraphics_CHI21} and 3D-printed artifacts~\cite{Holloway_3DPrintedMaps_ASSETS19, Holloway_AccessibleMapsForTheBlind-Comparing3DModels_CHI18, Holloway_TactIcons_CHI23, Crawford_CoDesigning3DPrintedTactileMap_ASSETS24} to refreshable tactile displays~\cite{Ivanchev_PreJourneyVisOfTravelRefreshable_SpecialNeeds2014, Reinders_RefreshableTactileDisplaysMeetConversationalAgents_TVCG25} and dynamic haptics using vibromotors~\cite{Yatani_SpaceSenseHapticFeedback_CHI12, Hong_HapticFeedbackForPathTracing_ASSETS17} or force-feedback input controllers~\cite{Tornil_ForceFeedbackPointing_UA04}. Despite significant advancements, prior work has not explored streetscape tool accessibility, our focus.

Beyond traditional maps, researchers have investigated real-world navigation support~\cite{Williams_PrayBeforeYouStepOut_ASSETS13, Saha_LastFewMetersWayfindingsForPeopleWithVI_ASSETS2019, Kameswaran_NavTechUseByBLV_ASSETS20, Giudice2008}, such as maintaining orientation during travel~\cite{Giudice2008}, navigating the ``last few meters'' to a destination~\cite{Saha_LastFewMetersWayfindingsForPeopleWithVI_ASSETS2019}, or avoiding obstacles~\cite{Williams_PrayBeforeYouStepOut_ASSETS13}. Others have developed and studied custom guidance systems such as \textit{SoundScape}~\cite{Microsoft_Soundscape_2023}, \textit{FootNotes}~\cite{Gleason_FootNotesGeoReferencedAudioAnnotations_IMWUT18}, and \textit{NavCog}~\cite{Ahmetovic_NavCog_MobileHCI16, Sato_NavCog3_TACCESS19}---all which combine real-time location sensing with spatial audio and/or audio guidance. Much of this research has matured into popular commercial tools or features, such as \textit{``Voice Guidance''} in Google Maps~\cite{Google_GoogleMapsVoiceGuidance_2023} or beacon-based audio navigation in \textit{BlindSquare}~\cite{BlindSquare}, \textit{GoodMaps}~\cite{GoodMaps}, and \textit{VoiceVista}~\cite{VoiceVista}. While \sysname is not an \textit{in situ} navigation tool, it is designed to support route planning and open exploration before travel---in ways not previously possible (\textit{e.g.,} finding business entrances \textit{a priori})---and should synergistically support these \textit{in situ} tools.

\sysname integrates real-time AI to support image-based analysis of street scenes and geographic queries, which connects to emerging work combining geographic information systems (GIS) with LLMs. For example, \textit{ChatGeoAI}~\cite{Mansourian_ChatGeoAI_IJGI24} and \textit{GIS Copilot}~\cite{Akinboyewa_GISCopilot_ArXiv2024} enable novice analysts to perform complex geospatial operations using LLMs. Relatedly, though not backed by LLMs, prior work has explored making geographic-based query and visualization systems accessible \cite{Thomas_Atlas_W4A08,Li_AltGeoViz_VIS24, Sharif_UnderstandingGeospatialDataVisForScreenReaderUsers_ASSETS22,Zhao_DataSonificationForVI_TOCHI08}. For example, \textit{Atlas.txt} introduced a novel data-to-text natural language generation system to communicate geo-referenced information through screen readers; however, it did not support interactive explorations. Most relevant to our work, \textit{VoxLens} \cite{Sharif_UnderstandingGeospatialDataVisForScreenReaderUsers_ASSETS22} enables screen reader users to verbally query geospatial data---for example, to understand COVID-19 cases by US state---and Reinders \textit{et al.'s} recent Wizard-of-Oz (WoZ) study examining refreshable displays with WoZ'd conversational agents~\cite{Reinders_RefreshableTactileDisplaysMeetConversationalAgents_TVCG25}. In both works, the authors examine and taxonomize the types of questions and information BLV users are interested in with accessible geoanalytics. We pursue a similar analysis of user questions but applied to streetscape imagery. Moreover, while promising, the above tools are focused on traditional 2D map-based representations rather than the integration of geospatial data, immersive imagery, and virtual navigation as we are here.

\begin{figure*}[tb] 
  \centering 
  \includegraphics[width=\textwidth]{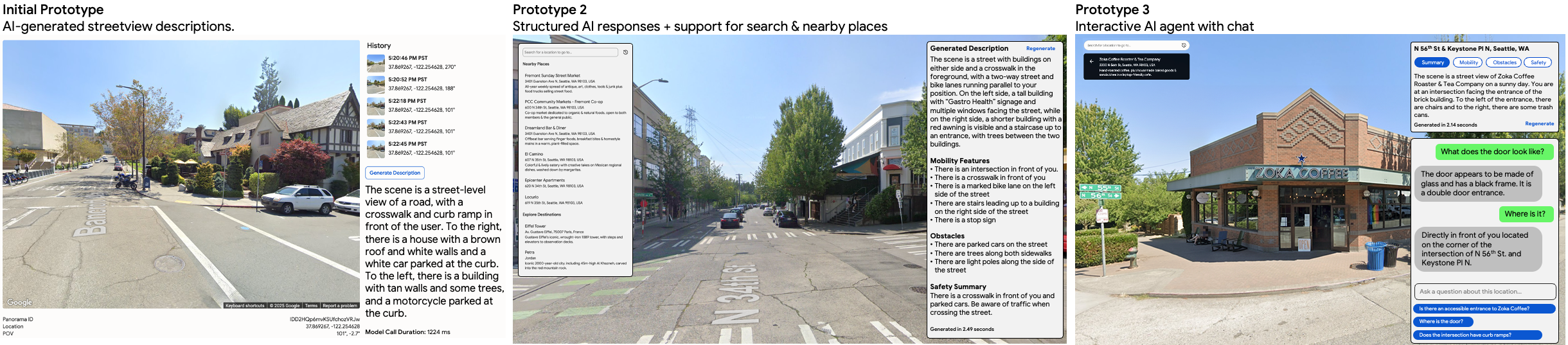}
  \vspace{-2.0em} 
  \caption{We iteratively designed \sysname with two blind collaborators, improving the navigation experience, integrating additional geographic knowledge bases (\textit{e.g.,} nearby places, intersections, \textit{etc.}), adding in an AI chat interface, implementing additional hot key support, and refining the AI prompts. For legibility, the on-screen text size was artificially increased in these screenshots---text is natively voiced by \sysname or read by the screen reader.}
  \Description{Shows three screenshots of early prototypes progressively evolving from left-to-right. (1) On the left screenshot, we should Prototype 1, which shows an interactive Google Street View (GSV), the user's history of views, and an AI-generated description for the current view. It also shows a button to generate another description and some metadata about the GSV pano. (2) In Prototype 2, we added in more integration with Google Maps like Google Places search and improved our AI prompts and output. Moreover, we started cleaning up the UI to look and feel more like regular GSV but with accessibility built in; (3) Finally, in Prototype 3, we added our initial AI Chat agent. The UI shows a chat between the agent and human as well as suggested questions at the bottom.}
  \label{fig:initial_prototypes}
\end{figure*}

\subsection{Accessible XR and Virtual Environments}
\sysname does not strictly fit into traditional definitions of Augmented Reality (AR) or Virtual Reality (VR)~\cite{Speicher_WhatIsMixedReality_CHI19} but its immersive, first-person perspective and interactive navigation share characteristics with both, and we draw on emerging mixed-reality (XR) accessibility guidelines ~\cite{Mott_AccessByDesignForVR_ISMAR19, Dudley_InclusiveImmersion_VR23}. Towards AR accessibility, research focuses on enhancing access to the \textit{physical world} through interactive computing, including via sensory substitution techniques~\cite{Danilov_Brainport_JIN05, Bach_SeeingWithTheBrain_IJHCI03}, sonification~\cite{Wilson_SWAN_ISWC07, Ahmetovic_SonificationOfRotation_PerCom19, May_SpotlightsAndSoundscape_TOCHI20}, and, most recently, real-time object recognition and description using smartphone cameras~\cite{Seeing_AI, Be_My_AI, Chheda_ArtInsight_IUI25, Chang_WorldScribe_UIST24, Gonzalez_AIPoweredScenceDesc_CHI24, apple_live_descriptions_2024} or AR glasses~\cite{Zhao_DesigningARVisForStair_UIST19, Lee_AIPoweredARSports_ISMAR24, Lee_CookAR_UIST24, Granquist_OrCam_JVIB21}. For example, \textit{WorldScribe}~\cite{Chang_WorldScribe_UIST24} uses emerging vision-language models (VLMs) to provide customized, succinct descriptions of a BLV's user's smartphone camera view. While related, \sysname differs by operating on pre-captured 360° imagery rather than live sensor data. Moreover, \sysname's interaction experience is disembodied, which introduces unique HCI challenges related to virtual locomotion, navigation, and views---while trying to maintain a spatial mental model of the world.

In this way, \sysname is most related to research in accessible VR, which attempts to make virtual environments perceivable and navigable without vision, including via accessible input mechanisms like instrumented canes~\cite{Zhao_EnablingPeopleWithVIToNavigateVR_CHI18, Siu_VRWithoutVision_CHI20} or echolocation~\cite{Andrade_EchoHouse_OzCHI18} as well as output modalities such as spatialized audio and haptic feedback~\cite{Walker_UsingVEToPrototypeAuditoryNav_2005, Schloerb_BlindAid_IEEEHaptics10, Wedoff_VirtualShowdownAccessibleVR_CHI19}. One early influential work includes \textit{BlindAid}~\cite{Schloerb_BlindAid_IEEEHaptics10}, a VR system using haptics and spatial audio to help BLV users learn layouts for Orientation and Mobility (O\&M) training, sharing \sysname's goal of virtual exploration for environmental understanding. More recently, researchers have applied real-time AI techniques in VR to improve accessibility~\cite{Chen_EnvisionVR_ArXiv25, Collins_AnAIGuideToEnhanceSocialVR_ASSETS24, Zhao_SeeingVR_CHI19}, much as we do with \sysname. For example, \textit{EnVisionVR}~\cite{Chen_EnvisionVR_ArXiv25} uses a VLM to support real-time scene descriptions and virtual object interaction; however, this is with purely \textit{virtual} scenes rather than physical world data projected into an immersive environment.

Finally, we draw on the long history of accessible 3D first-person video games such as \textit{Shades of Doom}~\cite{GMA_ShadesOfDoom}, \textit{AudioQuake}~\cite{Atkinson_AudioQuake_Sandbox06}, and \textit{Terraformers}~\cite{Westin_Terraformers_ICDVRAT04}, which demonstrate the feasibility of complex 3D navigation and interaction using keyboard interaction and spatialized sound. While \sysname does not currently include spatialized audio, it borrows keyboard input mechanics for panning and movement common in these games (\textit{e.g.,} arrows for movement).



\subsection{Image Accessibility and AI Descriptions}
Streetscape tools are fundamentally reliant on visual imagery, posing a significant accessibility barrier. Traditionally, image accessibility has depended on manually authored alternative text (alt-text)~\cite{W3C_WAI_Images}. However, alt-text is frequently missing~\cite{Gleason_ImageDescFailed_WWW19, Williams_SupportingQualityAltText_W4A22}, often inadequate for conveying the rich detail of complex scenes like streetscapes, and impractical to manually create for the billions of images hosted by services like GSV and Mapillary~\cite{Google_StreetView_Stats, Mapillary_Stats}. This necessitates automated approaches. Recent advances in AI, particularly LLMs and VLMs, have enabled powerful new image description techniques~\cite{Wu_AutomaticAltText_CSCW2017, Gubbi_ContextAwareImageDesc_ASSETS24, Dognin_ImageCaptioningLessonsLearnedFromVizWiz_JAIR20, Lee_ImageExplorer_CHI22, Zhao_AIVision-IMWUT2024} and can support \textit{Visual Question Answering} (VQA) systems where users ask specific questions about images~\cite{Antol_VisualQuestionAnswering_ICCV15}. These technologies are increasingly integrated into mainstream platforms---both Apple \textit{VoiceOver}~\cite{Apple_VoiceOverRecognition_2024} and Android \textit{TalkBack}~\cite{Google_TalkBackImageDesc_2024} support AI-generated image descriptions---as well as tools like \textit{BeMyAI}~\cite{Be_My_AI} and \textit{Seeing AI}~\cite{Seeing_AI}. While important progress, these AI-based approaches are built to support general image-based descriptions and lack geographic context, viewport history, and support for virtual inquiry, as we do with \sysname.

Recognizing the limitations of static alt-text descriptions, others have also explored interactive approaches such as \textit{ImageExplorer}~\cite{Lee_ImageExplorer_CHI22} and \textit{AI Vision}~\cite{Zhao_AIVision-IMWUT2024}, which provide hierarchical descriptions and detailed object exploration \textit{within} an image. While enhancing interaction with a single image, these systems are not designed for navigating \textit{between} immersive scenes or grounding interactions within a dynamically changing virtual viewpoint. \sysname advances this area by uniquely integrating context-aware AI capabilities directly within a dynamic, navigable, 3D spatial interface.

\subsection{Street Scene Analysis}
Finally, \sysname's ability to provide meaningful descriptions relies on advancements in AI-based street scene analysis and understanding~\cite{Cordts_Cityscapes_CVPR16, Neuhold_MapillaryVistasDataForSemantic_ICCV17, Zhou_SceneParsingThroughADE20K_CVPR17, Chen_DeepLab_TPAMI18}. Most relevant to our work is research in automatically identifying pedestrian-related and accessibility features, including curb ramps~\cite{Hara_Tohme_UIST14, Hara_AutomaticCurbRampDetection_HCOMP2013}, sidewalk presence or condition~\cite{Weld_DeepLearningDetectSidewalkAccess_ASSETS19, Hosseini_GlobalScaleCrowdAITechniquesSidewalks_ArXiv22}, specific accessibility obstacles~\cite{Duan_ScalingCrowdAISidewalkAssessment_ASSETS22}, sidewalk surface materials~\cite{Hosseini_CitySurfaces_SCS22}, traffic signs~\cite{Balali_DetectionClassificationOfUSTrafficSignsFromGSV_ViE2015}, and assessing overall streetscape walkability ~\cite{Jeon_DeepLearningStreetPanoramaWalkability_LUP23, Lee_ComputerVisionStudyOfStreetScapePedestrian_Sustainability22}. Perhaps most related is \textit{BusStopCV}~\cite{Kulkarni_BusStopCV_ASSETS23}, a web app with an embedded \textit{YOLOv8} model~\cite{Redmon_YOLO_ArXiv16} that analyzes the user's streetscape view in real-time to help identify and collect accessibility information about bus stops (\textit{e.g.,} presence of benches, shelters). Though not aimed at screen reader accessibility, BusStopCV demonstrates the potential of automatically extracting information from GSV in real-time as the user interacts with a street view scene.

Together, the above work demonstrates the power of AI-based analysis of street view imagery at scale; however, the focus is either on CV model improvements or efficient data acquisition for urban planning and mapping databases (\textit{e.g.,} ~\cite{Hosseini_GlobalScaleCrowdAITechniquesSidewalks_ArXiv22}). In contrast, our goal is to leverage these emerging techniques to provide real-time, interactive, and conversational access to the visual information within streetscapes. We do not aim to advance the state-of-the-art in CV models themselves but rather to apply these continuously improving models to create a novel accessible mapping experience.

\section{Designing \sysname}
To design and build \sysname, we followed an iterative, human-centered design process that included nine co-design sessions with our two blind collaborators (CB1, CB2)---\revised{who are also coauthors, feedback from professional designers and engineers with expertise in maps and AI-based mixed-reality, and drawing on our own decades of experience in creating accessible tools}. Co-design sessions began with initial ideation and use of an early-stage prototype, which was progressively improved through feedback (\autoref{fig:initial_prototypes}). We synthesize key findings and design considerations below.


\textbf{Initial reactions.} Though both of our blind collaborators were aware of streetscape tools, they had never used them due to inaccessibility. As CB2 described, \myquote{I have heard of [Street View] but I couldn't use it since it is not accessible.} After using the \sysname prototype for the first time, CB1 responded \myquote{I think there is massive value here} and CB2 said, \myquote{This is so exciting, just having a conversation about the street [with AI] is really amazing.} Both co-designers emphasized the importance of user agency and freedom---the ability to explore, encounter features of interest, and drill down: \myquote{I just want to learn more about what is around. I want to explore. And part of exploration is serendipity and the ability to see something and investigate more} (CB1).


\textbf{Learning mental models of street view.} After some initial sessions, we realized that the concept of street view itself was unclear; thus, we worked to improve our explanation of \textit{what} street view is (the concept of immersive and fully interactive 360° imagery), \textit{how}, \textit{where}, and \textit{when} the images are taken (from the street by instrumented vehicles with 360° cameras), and \textit{how} the immersive images are distributed and connected (available every 10-15 meters). Moreover, the inherent limitations of street view were unclear: how \textit{close} could you get to buildings? Could you go \textit{inside} buildings? 


\textbf{Navigation.} For navigation, both co-designers emphasized wanting pertinent information on every ``virtual'' step (movement from one pano to another), including street addresses, nearby places, distance moved, and whether they had previously visited a location. Both worried, however, about verbosity, \myquote{The information is really great but the verbosity will have to be trimmed down}, and suggested allowing the user to control what they wanted to hear and adding non-verbal auditory signals (\textit{i.e.,} earcons) for certain events.  Moreover, both suggested additional movement controls beyond heading changes and virtual steps. For example, when virtually navigating from his apartment to a local shopping mall with \sysname---along a route he knows well---CB1 suggested adding a ``jump'' feature to move by some larger increment or to a proximal landmark along the street (\textit{e.g.,} an intersection).

\textbf{Orientation.} Orienting oneself geographically while maintaining notions of heading, relative position, and nearby places can be challenging and becomes even more complicated with virtual movement (where actions are disembodied). Our co-designers suggested adding quick access to key shortcuts for \myquote{Where am I?}, \myquote{What am I facing?}, \myquote{What road am I on?}, and \myquote{What is the next intersection?}. 


\textbf{Use of AI.} The AI chat interface seemed to gain increasing importance as we iterated on the tool. As CB2 stated, \myquote{The chat tool is so important; once you get to a destination, you can learn so many things like, how busy the street is? What are the nearby intersections? What does the door look like? Where is the door?}. Beyond conversations about the scene, our co-designers suggested supporting commands to use the AI to direct movement or view changes: \myquote{Can you ask the AI chat to move you or turn you around? Or to take you places: 'take me to the nearest ice cream shop?} Finally, while we initially auto-generated an AI description upon every view change (either from panning or movement), to reduce information overload, we ultimately decided to require an explicit hotkey press for AI descriptions (\keys{Alt} + \keys{D}) and the AI chat (\keys{Alt} + \keys{C}).

\subsection{Design Considerations}
Drawing on our co-design sessions and key literature, we synthesized the following design considerations:

\textbf{Concision.} Between the street view image itself and the surrounding geography (\textit{e.g.,} roads, intersections, nearby places), the streetscape navigation experience is information dense. A key design challenge then is providing high value, pertinent information with every viewpoint change: concise and clear.

\textbf{Support conversational exploration.} Allow the user to easily engage with the AI chat to quickly ask questions and follow-ups about the scene and location. Chat questions are inherently personalized to the user's interest in the moment: \myquote{Does the sidewalk look shaded?}, \myquote{Is the entrance to the coffee shop wheelchair accessible?}, \myquote{Is there anything surprising along this route?}.

\textbf{Leverage streetscape images.} At every interaction, harvest as much as possible from the street view image relevant to a blind or low-vision pedestrian, including safety features, sidewalk obstacles, and ambient qualities like shade. Go beyond what is currently available in existing map APIs (\textit{e.g.,} in Google Maps or OpenStreetMaps).

\textbf{Support ``last 10 meters'' navigation.} Traditional mapping tools support origin-to-destination routing but only at the building-level granularity. We aim to support ``last 10 meters'' navigation~\cite{Saha_LastFewMetersWayfindingsForPeopleWithVI_ASSETS2019}, including the accessibility of sidewalks in front of buildings and the location of entrances.

\textbf{Enable personalization.} Allow users to tell the AI about themselves and their needs (\textit{e.g.,} mobility aid usage, preferred verbosity level, types of points-of-interests).

\textbf{Key use cases.} Finally, we distilled the following key use cases: (1) point-of-interest (POI) investigation---\myquote{What does a destination look like? Where is the entrance?} (2) open-world navigation---\myquote{I want the freedom to explore a region}; (3) route planning---\myquote{What might my walking route look like from \textit{A} to \textit{B}?}; and virtual tourism---\myquote{I've always wanted to visit Tokyo, Sydney, or even the Grand Canyon.} 





\section{The \sysname System}
\begin{figure*}[tb] 
  \centering 
  \includegraphics[width=\textwidth]{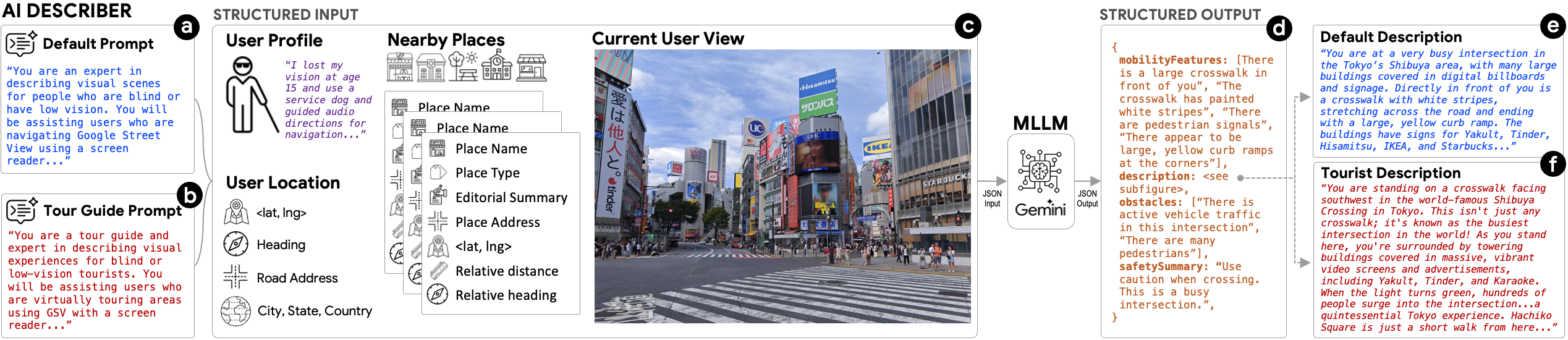}
  \vspace{-2.0em} 
  \caption{\revised{\textit{AI Describer} uses one of two prompts: (a) a "default prompt" emphasizing navigation and safety for BLV pedestrians and (b) a "tour guide" prompt that additionally emphasizes tourism information such as historic and architectural context. (c) The MLLM model is fed the custom prompt along with an optional user profile and dynamic, context-aware input such as the user's location, nearby places, and their current view. The output is structured with (d) mobility features, obstacles, and a safety summary along with the (e) default or (f) tourist description, tailored from the prompt personality.}}
  \Description{The AI Describer system diagram showing the two prompts, the structured input, and the structured output.}
  \label{fig:ai_describe_diagram}
\end{figure*}

Informed by our co-design sessions and prior literature, we built and evaluated the final \sysname prototype, the first interactive streetscape mapping tool for BLV users, which uses a context-aware, multimodal AI to describe scenes (\autoref{fig:teaser}) and accessible navigation controls. With \sysname, users can virtually navigate the physical world via keyboard interactions, hear AI descriptions of the current scene, or chat with a context-aware, multimodal AI agent. See the video demo. Below, we describe key components, including panning and movement, the AI subsystem, and other accessible controls. We built the app to be self-voicing, though this can be toggled on/off to work with screen readers. For the latter, we use ARIA live regions (HTML elements with \texttt{aria-live}) to provide status messages and AI output to a screen reader. 

\begin{table}[b]
\centering
\footnotesize
\renewcommand{\arraystretch}{1.1}
\setlength{\tabcolsep}{3pt}
\begin{tabular}{@{}lp{0.8\columnwidth}@{}}
\textbf{Key} & \textbf{Action} \\
\midrule
\multicolumn{2}{@{}l}{\textbf{Movement Controls}} \\
\rowcolor[HTML]{EFF2F6} 
\keys{{$\leftarrow$}}~~\keys{{$\rightarrow$}} & Rotate left or right 45° \\
\rowcolor[HTML]{FFFFFF} 
\keys{{$\uparrow$}}~~\keys{{$\downarrow$}} & Move forward/backward at current heading (if possible) \\
\rowcolor[HTML]{EFF2F6} 
\keys{Alt} + \keys{B} & Go \textbf{\underline{b}}ack to the last location (pano) \\
\rowcolor[HTML]{FFFFFF} 
\keys{Alt} + \keys{J} & \textbf{\underline{J}}ump to the next intersection or 70 meters (230 feet), whichever is first at your current heading \\
\multicolumn{2}{@{}l}{\textbf{AI Interaction}} \\
\rowcolor[HTML]{EFF2F6} 
\keys{Alt} + \keys{D} & \textbf{\underline{D}}escribe the current view with AI \\
\rowcolor[HTML]{FFFFFF} 
\keys{Alt} + \keys{C} & \textbf{\underline{C}}hat with the AI agent (typing) \\
\rowcolor[HTML]{EFF2F6} 
\keys{Alt} + \keys{Spacebar} & Talk with the AI agent (speaking) \\
\rowcolor[HTML]{FFFFFF} 
\keys{Alt} + \keys{A} & Repeat the previous output \textbf{\underline{a}}gain \\
\rowcolor[HTML]{EFF2F6} 
\keys{Esc} & Stop the current speech output \\
\multicolumn{2}{@{}l}{\textbf{Location Information}} \\
\rowcolor[HTML]{EFF2F6} 
\keys{Alt} + \keys{W} & \textbf{\underline{W}}here am I? Get current address and heading \\
\rowcolor[HTML]{FFFFFF} 
\keys{Alt} + \keys{N} & Get information about \textbf{\underline{n}}earby places \\
\rowcolor[HTML]{EFF2F6} 
\keys{Alt} + \keys{I} & Get information about the current and next \textbf{\underline{i}}ntersection (if any) \\
\rowcolor[HTML]{FFFFFF} 
\keys{Alt} + \keys{M} & Get possible \textbf{\underline{m}}ovements at current location \\
\multicolumn{2}{@{}l}{\textbf{Meta Information}} \\
\rowcolor[HTML]{EFF2F6} 
\keys{Alt} + \keys{V} & Get your \textbf{\underline{v}}isit history for current pano \\
\rowcolor[HTML]{FFFFFF} 
\keys{Alt} + \keys{P} & Get the date and photographer of current pano (\textbf{\underline{p}}hotograph) \\
\end{tabular}
\caption{List of hotkeys and their associated behaviors.}
\label{tab:hotkeys}
\end{table}

\subsection{Panning and Movement}

We describe the user actions and underlying algorithms to support keyboard-based, non-visual interaction with \sysname. As they interact, the user can trigger an AI description of their current view with \keys{Alt} + \keys{D} or chat with the AI agent via \keys{Alt} + \keys{C} (typing) or \keys{Alt} + \keys{Spacebar} (speaking). For example, after turning 90° or taking a step, the user can hit \keys{Alt} + \keys{D} to hear an AI-generated summary of the current view. Specific thresholds mentioned below were developed iteratively through internal testing and our co-design sessions (but are parameterized and easily changeable).

\textbf{Panning.} Unlike traditional 2D images, street view images are high-resolution panoramas---140 megapixels or 16,733 x 8,366 pixels \cite{KRDO_Google_StreetView_Camera}---spherically projected into an immersive, interactive scene (similar to a VR world). The GSV API provides decimal-level control of the user's point-of-view (POV) with 360° horizontal and 180° vertical panning. To simplify interaction and limit disorientation, we discretize the horizontal (heading) panning space into octants (45° increments) and fix the vertical panning (pitch) to 0° (so, the user is looking ``straight out'' roughly at eye level). The user can pan left and right 45° using the \keys{{$\leftarrow$}}~~\keys{{$\rightarrow$}} arrow keys, respectively. 

As the user pans, \sysname immediately voices the current heading as a cardinal or intercardinal direction (\textit{e.g.,} \myquote{Now facing: North} or \myquote{Northeast}), expresses whether the user can move forward along that heading and, if so, to what road address, and also explains whether the user is now facing a nearby place. To generate the places description, we maintain a list of nearby places at each panorama (pano) location using the \textit{Google Places API}~\cite{google_places_api}. When the user's heading shifts, we describe places in front of the user---within a 45° angle of the current heading and a maximum distance threshold of 35 meters.   

\textbf{Taking a step.} If a nearby pano is available at the current heading, the user can take a virtual ``step'' forward using the \keys{{$\uparrow$}} arrow. Similarly, if a pano is available opposite their current heading (180°), they can step backwards with \keys{{$\downarrow$}} (but will still maintain their current forward heading). These steps are approximately 5-15 meters depending on the GSV pano distribution in that geographic region. 

\begin{figure}[tb] 
  \centering 
  \includegraphics[width=\columnwidth]{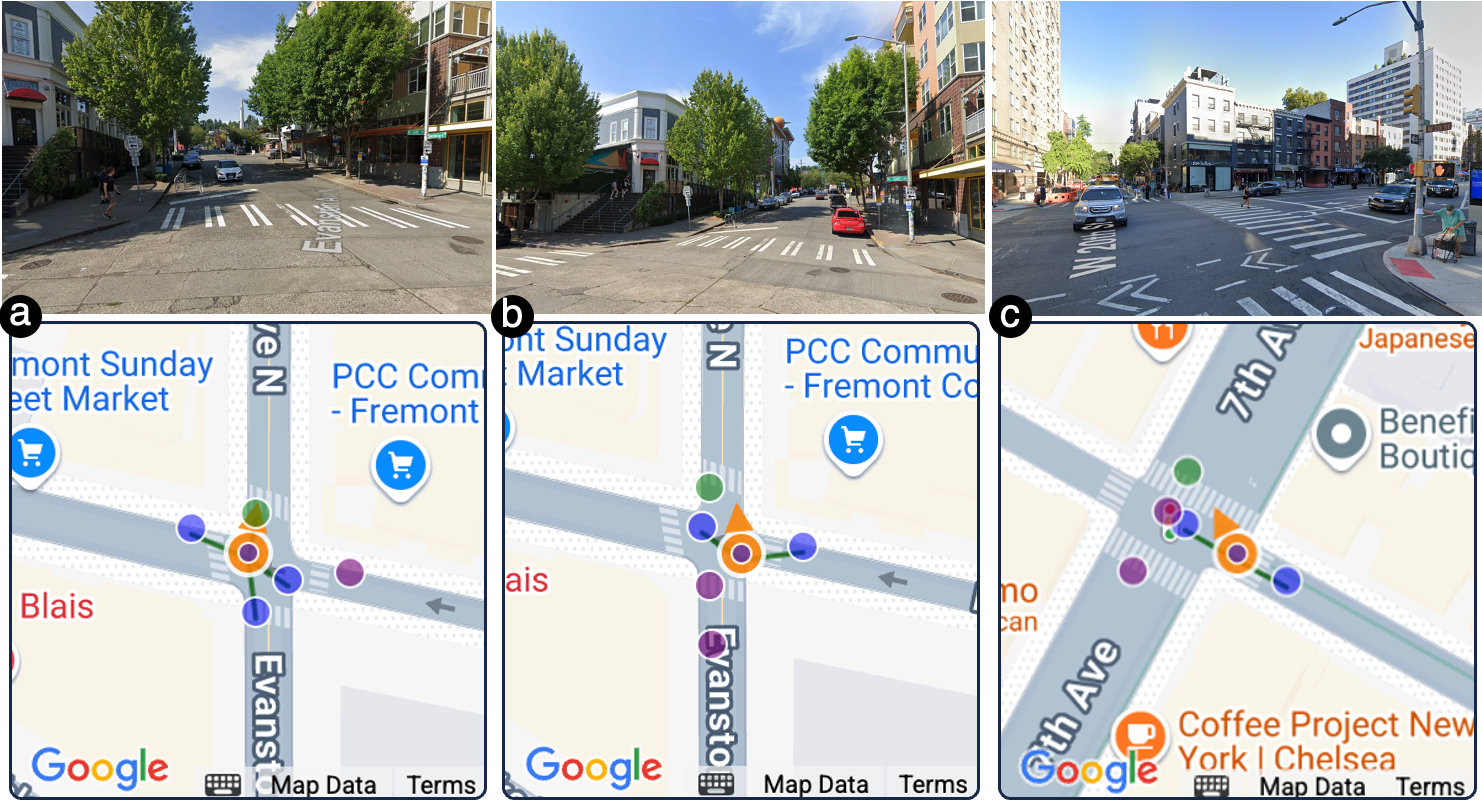} 
  \vspace{-2.0em}
  \caption{To support accessible streetscape navigation, we construct a custom egocentric GSV graph. From the user's current location (orange circle with heading), connected \texttt{StreetViewLinks} are drawn with edges and shown in blue; the current ``move forward'' pano is shown in green. (a) In this intersection, the built-in GSV graph is fine---the user can appropriately move north, south, east, and west; (b) In contrast, here there are only two built-in links restricting travel to east-west; our custom graph addresses this limitation and still allows for north, south, east, west travel, matching the intersection. (c) Similarly, this four-way intersection only has two built-in GSV links (two directions of travel) while our solution allows for all four, again matching the roads.}
  \Description{The image shows three subfigures, each intended to show weaknesses in GSV's default node-link network connecting navigable panos. (1) In subfigure A, the built-in GSV graph for this intersection is fine: the user can move in all four directions as expected along the North-South and East-West roads; (2) In subfigure B, however, the built in GSV graph only allows for two East-West movements as if the North-South road does not exist. This is confusing to users. In contrast, our solution properly allows all four movements. (3) Finally, in subfigure C, we again show a four-way intersection where GSV's built-in network only allows for movements in two directions rather than all four along the roadways.}
  \label{fig:streetviewai-gsv-links} 
\end{figure}

To calculate navigable panos reachable by a ``step'', we derived our own custom algorithm suitable for non-visual interaction. Each \texttt{StreetViewPanorama} object in the GSV API maintains a list of connected panos; these \texttt{StreetViewLinks} include a heading (oriented from the current pano), a description, and a unique `pano id`. Thus, one can navigate in GSV simply by traversing this node-link network (drawn as blue circles and edges in \autoref{fig:streetviewai-gsv-links}); however, Google does not publish how they decide which panos to connect and, in practice, we found this network insufficient to enable non-visual navigation. For example, a user could be in the middle of a four-way intersection with a North-South street and an East-West street but at a pano with only two links, which artificially limits movement and causes confusion (see examples in \autoref{fig:streetviewai-gsv-links}). The problem is: the road network does not match the underlying GSV network.

Thus, for each location, we construct a custom egocentric GSV graph, which better matches user navigability expectations and the underlying road network. For this, we perform a parallel search of nearby panoramas along a 20x20 meter search grid with 5 meter step increments. We identify all panos within the grid but also all \texttt{StreetViewLink} panos (if not in the nearby list). For each found pano, we compute a distance and relative heading to the user's current location. With this data structure, we can immediately calculate the next and previous panos as the user changes their heading. If multiple panos exist along the same trajectory, we default to the closest one. If the user attempts to move forward or backward but no pano exists along that heading, we voice an \textit{available movements} status message (also triggerable by \keys{Alt} + \keys{M}), which expresses possible movements at the user's current location. For example, \myquote{You cannot move [forward/backward] along your current heading of East. You can move in three directions: Northwest, West, and Southeast.}

After taking a step, \sysname voices the movement type (\textit{e.g.,} step, jump), how far the user traveled, their current address, whether they moved to a new road, whether they have just arrived or left an intersection, a list of nearby places (within 50 meters) prioritized by direction, and whether they have previously visited the specific pano. For brevity, we only voice information that changes from the previous pano. Moreover, the list of nearby places is contextual and uses relative positions. For example, \myquote{Starbucks Coffee is now on your left 12 meters away} or \myquote{The Talking Book and Braille Library is still in front of you but now 32 meters away.}

\textbf{Jumping.} In addition to ``stepping'', the user can also \textit{jump} along their current heading by 70 meters (230 feet) or to the nearest intersection, whichever comes first. Because the Google Maps API does not provide road intersection data, we implemented our own detection algorithm. Our method casts a ray from the user's current location projected along their current heading and incrementally checks each point for an intersection with a specified step size (15 meters), search grid size (20x20 meters), and maximum distance (70 meters). To detect intersections at each step, we use the \textit{Google Roads API}~\cite{google_roads_api} to request roads and their geometry within the search grid. If two or more roads are found, we then run a parallel search for nearby panos within the same grid and determine if at least one pano has adjacent neighbors with assigned addresses of different roads. This works well in practice but does mean that bridges where two roads cross but do not intersect are treated the same as actual intersecting roads. After jumping, we generate a similar status message as a ``step'' but tell the user they ``jumped.''

\textbf{Teleporting.} Finally, the user can teleport by searching for and selecting a specific place or address in the ``search box'' field. This functionality is built on top of the \texttt{searchByText} endpoint in the Google Places API. As the user types, \sysname returns a list of matching results in real-time; the list includes a display name, address, location type, and brief editorial summary. Once a new destination is selected, we virtually ``teleport'' the user to the closest nearby pano. Crucially, to help orient the user upon landing, we automatically adjust the POV to point to the destination.

Once the user teleports, we generate a more verbose status message to orient the user to their new location, including announcing the city, state, and/or country if they changed. For example: \begin{myblockquote}
You teleported 2,393 km from the Acropolis of Athens to 38 Bankside in London, England. There are four places within 50 meters, including: a performing arts theater, Shakespeare's Globe, is ahead of you 26 meters away; a historical landmark, Sam Wanamaker Plaque and an event venue, Ingresso, are to your left less than 47 meters away, and a garden, Tate Community Garden, are to your right less than 49 meters away. You are facing South and can move in four directions: South, West, Northwest, and East.
\end{myblockquote}

\textbf{Go back.} If the user made a mistake or simply wants to return to their most recently visited pano, they can hit the \keys{Alt} + \keys{B} keys. This undo stack is currently of size one but is parameterizable for future iterations of the tool.

\begin{figure*}[!htbp]
    \centering
    \includegraphics[width=0.33\textwidth]{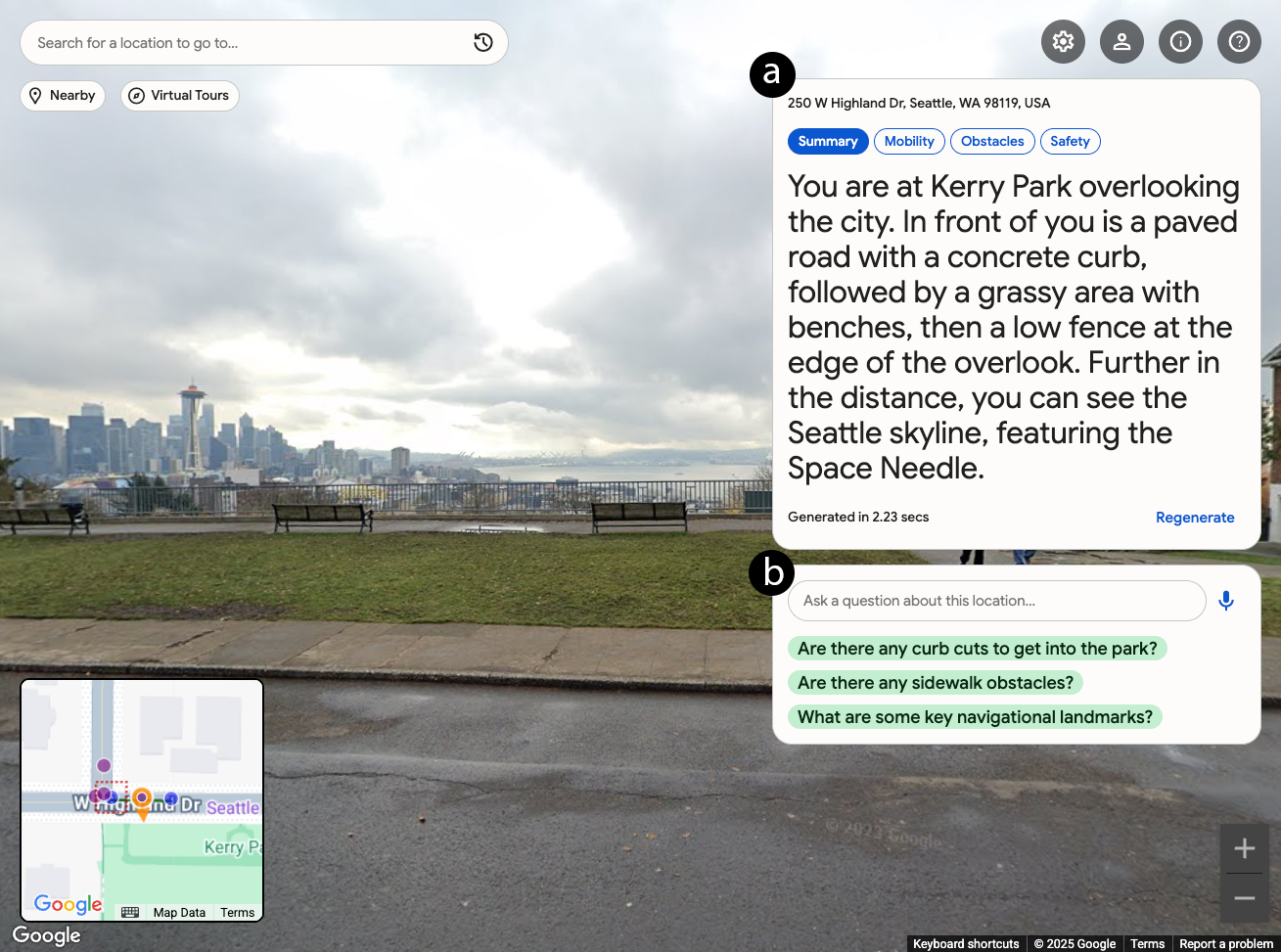}%
    \Description{An image showing Kerry Park overlooking the city of Seattle. The AI generated description (from AI Describe) shown on the right side of the StreetViewAI interface says: "You are at Kerry Park overlooking the city. In front of you is a paved road with a concrete curb, followed by a grassy area with benches, then a low fence at the edge of the overlook. Further in the distance, you can see the Seattle skyline, featuring the Space Needle."}%
    \includegraphics[width=0.33\textwidth]{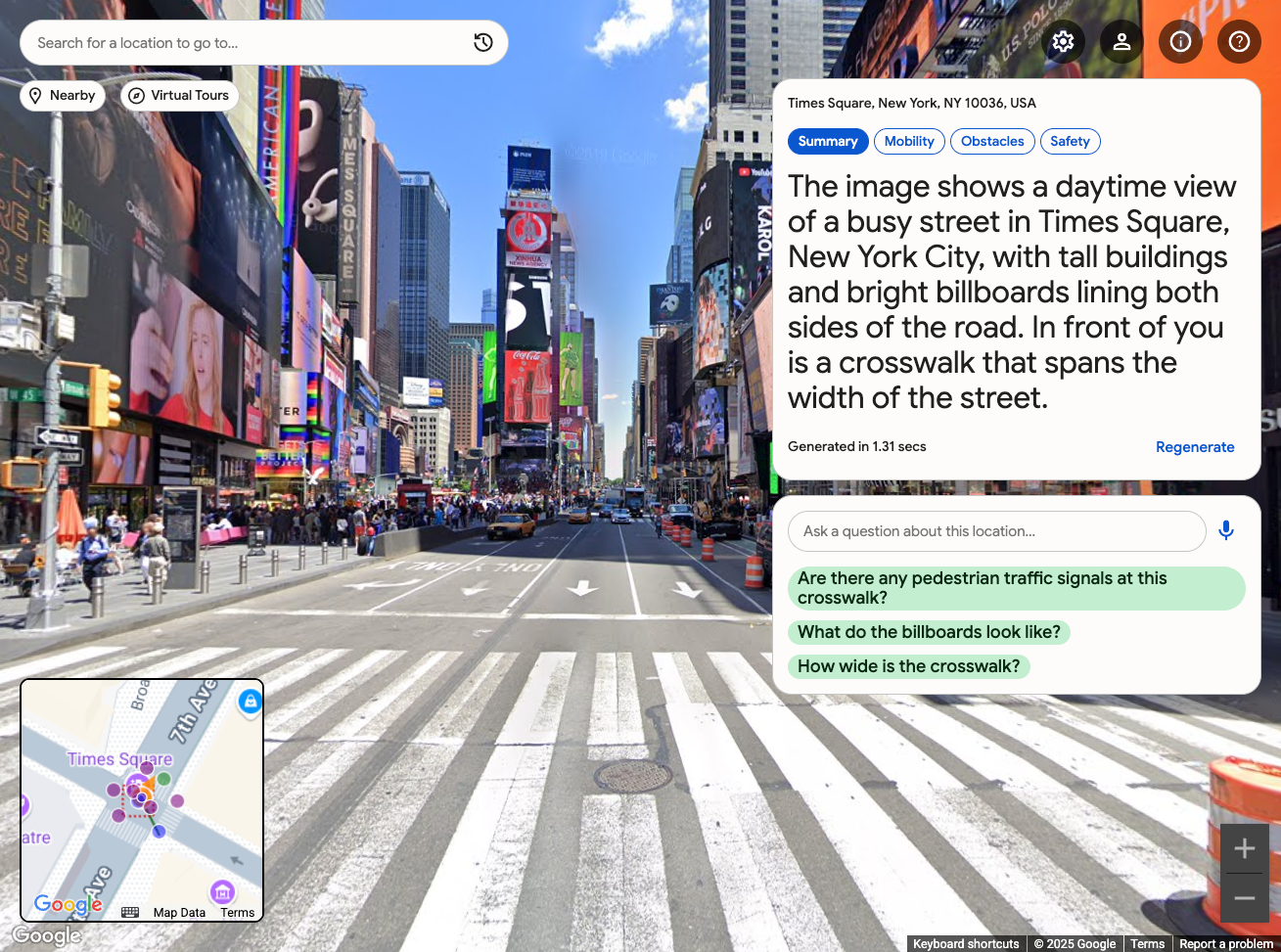}%
    \Description{An street view image of Times Square. The AI Describe generated description: "The image shows a daytime view of a busy street in Times Square, New York City, with tall buildings and bright billboards lining both sides of the road. In front of you is a crosswalk that spans the width of the street."}%
    \includegraphics[width=0.33\textwidth]{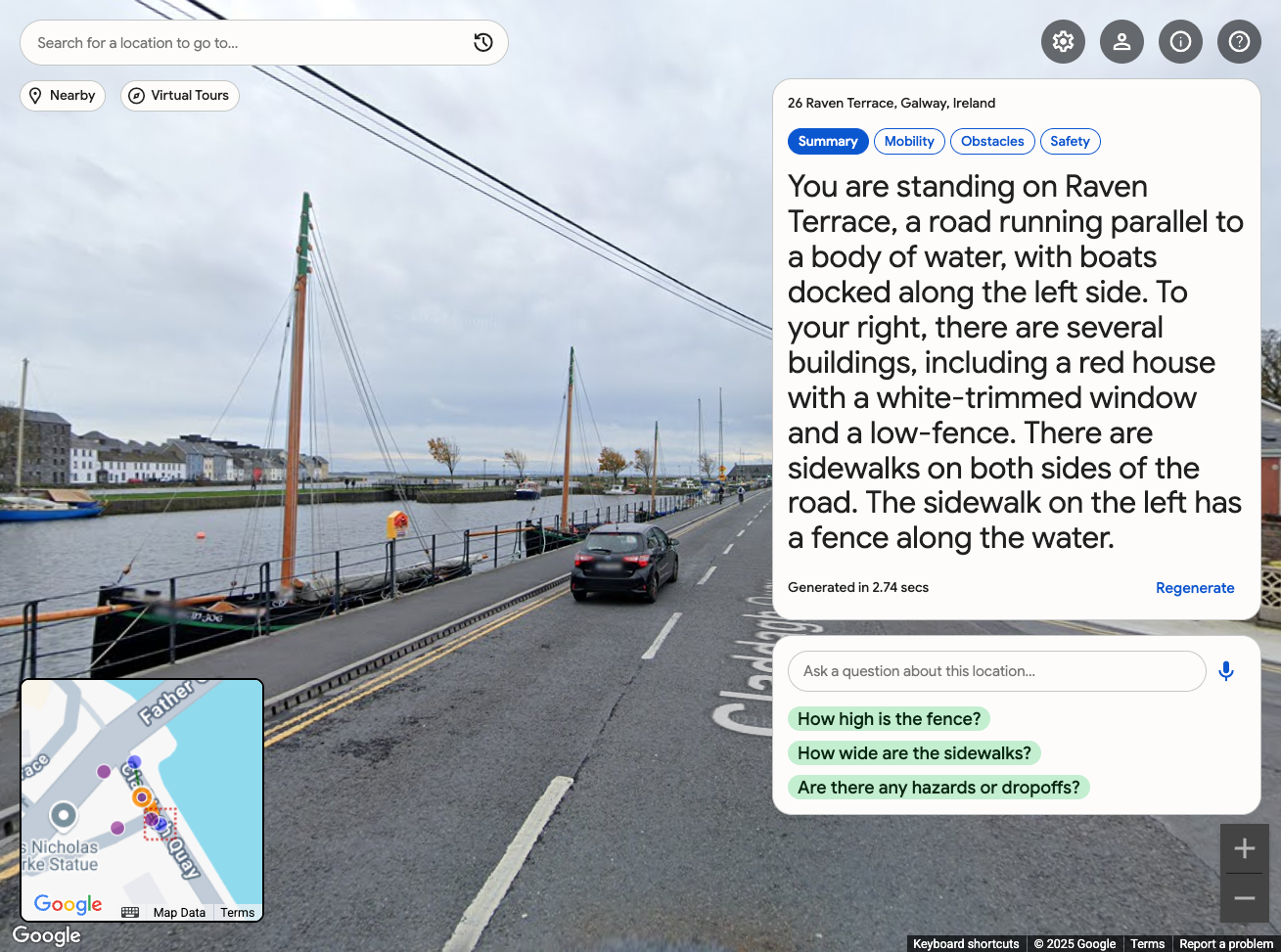}%
    \Description{A street scene in Galway, Ireland showing boats docked along the road. The AI Describe description: "You are standing on Raven Terrace, a road running parallel to a body of water, with boats docked along the left side. To your right, there are several buildings, including a red house with a white-trimmed window and a low-fence. There are sidewalks on both sides of the road. The sidewalk on the left has a fence along the water."}%
    
    \includegraphics[width=0.33\textwidth]{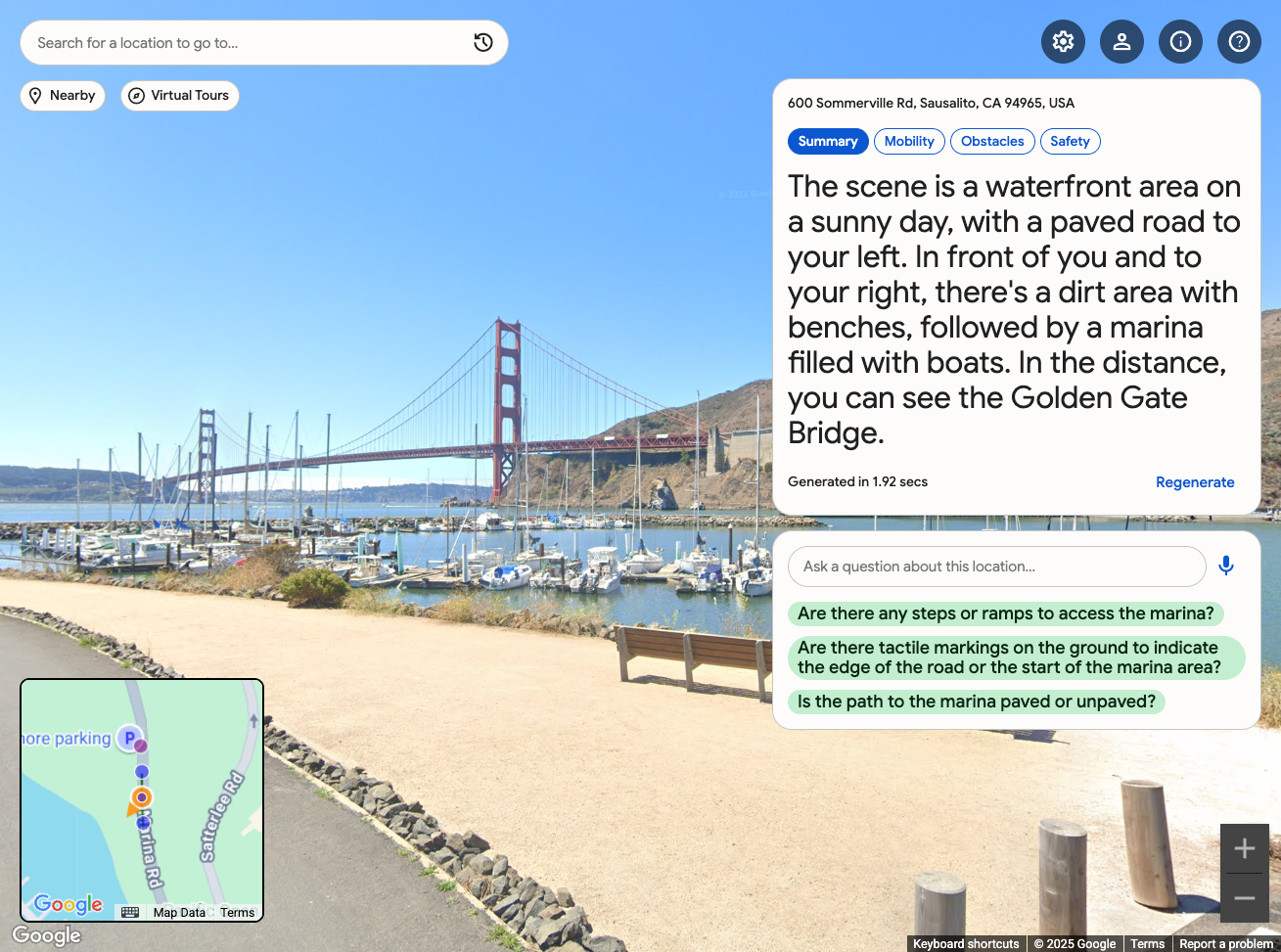}%
    \Description{The Golden Gate bridge as seen from Marina Rd in California. The AI Describe description is: "The scene is a waterfront area on a sunny day, with a paved road to your left. In front of you and to your right, there's a dirt area with benches, followed by a marina filled with boats. In the distance, you can see the Golden Gate Bridge."}%
    \includegraphics[width=0.33\textwidth]{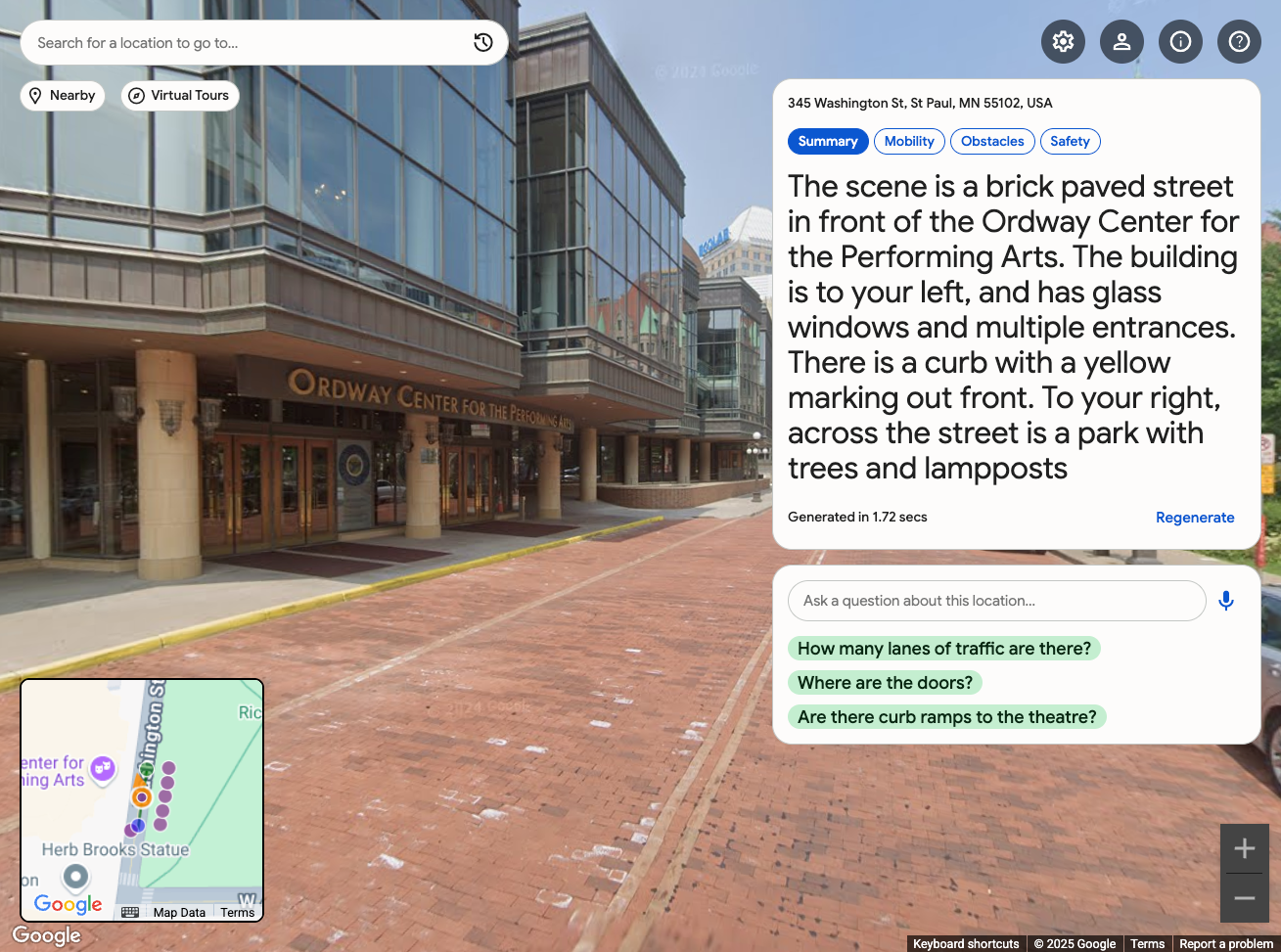}%
    \Description{ Outside the Ordway Theatre in St. Paul, MN. The AI Describe description is: "The scene is a brick paved street in front of the Ordway Center for the Performing Arts. The building is to your left, and has glass windows and multiple entrances. There is a curb with a yellow marking out front. To your right, across the street is a park with trees and lampposts."}%
    \includegraphics[width=0.33\textwidth]{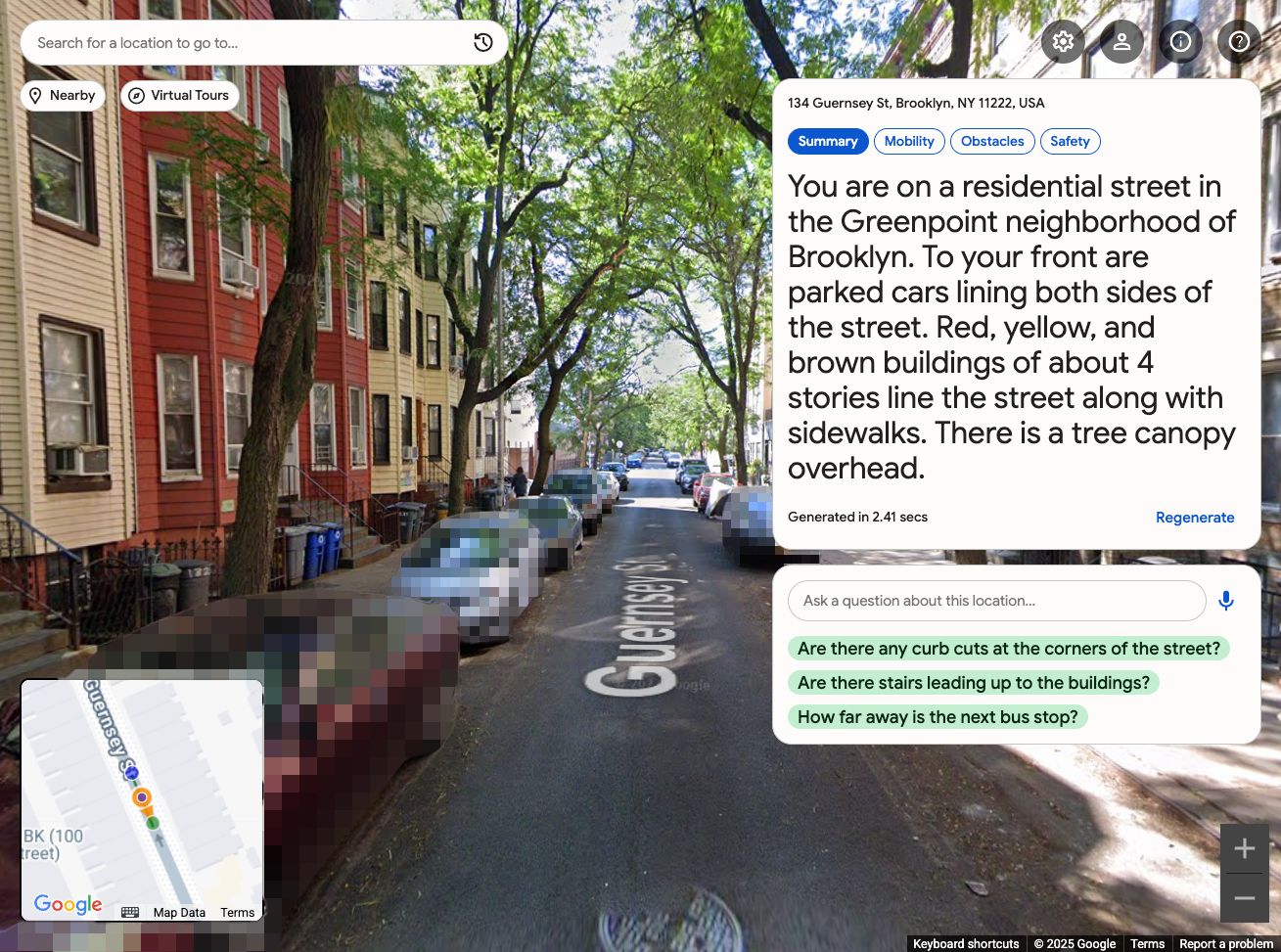}%
    \Description{A residential street in Brooklyn. The AI Describe description is: "You are on a residential street in the Greenpoint neighborhood of Brooklyn. To your front are parked cars lining both sides of the street. Red, yellow, and brown buildings of about 4 stories line the street along with sidewalks. There is a tree canopy overhead."}%
    \vspace{-1.0em}
    \caption{Example output from \sysname's \textit{AI Describer} across diverse urban and residential scenes. (a) The MLLM generated descriptions and (b) three context-relevant \textit{AI Chat} suggestions for followup questions, enabling serendipitous discovery.}
    \label{fig:ai_describer_examples}
\end{figure*}

\subsection{\sysname's Multimodal AI}
At any point, the user can trigger an AI-based analysis of their current view to learn more information. \sysname has three AI subsystems, all built on top of Google's \textit{Gemini Flash 2.0} model~\cite{google_vertex_gemini}, which was selected due to its benchmark performance, ease of prototyping, and multimodal support \cite{GoogleDeepMind_Gemini_2025}. The three AI subsystems include: \textit{AI Describer}, \textit{AI Chat}, and \textit{AI Tour Guide}, each which use custom prompts informed by prior work \cite{Chang_WorldScribe_UIST24, Chheda_ArtInsight_IUI25, Chheda_ChildrenArtwork_ASSETS24}, our co-design sessions, and iterated upon through internal testing. Unlike traditional AI-based image description tools \cite{Lee_ImageExplorer_CHI22, Wu_AutomaticAltText_CSCW2017, Seeing_AI}, our AI models are multimodal and can consider relevant contextual information about the user, surrounding geography, and the image itself (Figures \ref{fig:ai_chat_diagram}, \ref{fig:ai_describe_diagram}). We describe each input component before the AI subsystems.

\textbf{User profile}. Users can optionally set a user profile, describing their vision level, mobility needs, and other relevant information, which is included as context in the prompt. For example, \myquote{I lost my vision at age 19 due to Stargardt disease. For navigation, I use a service dog and guided audio directions in Google Maps. For travel, I prefer to walk, especially if the distances are under one mile.} If this profile is not set, the prompt asks the model to \myquote{assume the user is blind and may use a white cane or a guide dog for mobility.}

\textbf{Geographic context}. We also assemble a dynamic list of geographic elements nearby the user, including the user's currently selected place (if any), their closest address, current heading, neighborhood, city, state/province, country, and a sorted list of nearby places with their business names, types, editorial summaries, \texttt{<lat,lng>} positions, distances from the user, heading offsets, relative positions (\textit{e.g.,} left or behind the user).

\textbf{User's view}. Finally, we upload a 640x640 image of the user's current view, extracted from their current position and heading. Below, we describe each of the three AI subsystems. See supplementary materials for the full prompts.

\begin{table}[h!]
\centering
\footnotesize
\renewcommand{\arraystretch}{1} 
\setlength{\tabcolsep}{2pt}
\renewcommand{\tabularxcolumn}[1]{>{\raggedright\arraybackslash}p{#1}}
\begin{tabularx}{\columnwidth}{@{} l X X X @{}}
\toprule
\textbf{Feature} & \textbf{AI Describer} & \textbf{AI Chat Agent} & \textbf{AI Tour Guide} \\
\midrule
\rowcolor[HTML]{EFF2F6}
Primary Goal & Concise scene awareness for orientation \& navigation & Navigational cues, Interactive Q\&A, System control & Engaging, informative virtual tour experience \\
AI Persona & Expert scene describer for BLV users & Helpful, responsive conversational agent & Knowledgeable, engaging tour guide for BLV tourists \\
\rowcolor[HTML]{EFF2F6}
Key Focus & Key objects, spatial relationships, navigational elements & Answering questions, executing commands & Historical facts, cultural significance, points of interest \\
Output Style & Short (2-3 sentences) or JSON & Conversational text or speech & Descriptive narrative (4-5 sentences) \\
\rowcolor[HTML]{EFF2F6}
Interactivity & On-demand description generation & Real-time chat with conversation memory & On-demand enriched description \\
Full Prompt & Supplementary Material S1.A & Supplementary Material S1.B & Supplementary Material S1.C \\
\bottomrule
\end{tabularx}
\caption{Comparison of the three AI subsystems highlighting their distinct goals, personas, and functionalities.}
\label{tab:ai_subsystems_comparison}
\end{table}

\subsubsection{AI Describer}
To invoke the \textit{AI Describer}, the user can press \keys{Alt} + \keys{D}, which combines the three data components above within a custom prompt that begins: 

\begin{myblockquote}
You are an expert in describing visual scenes for people who are blind or have low vision. You will be assisting users who are navigating Google Street View using a screen reader. They rely on detailed audio descriptions to understand their surroundings. Your task is to describe this image concisely and accurately, capturing key details, as if you are describing it to a person who cannot see it.
\end{myblockquote} 
The prompt also tells the model to focus on eight specific areas, including key objects, spatial relationships, and navigational cues in the scene deemed relevant to a blind person and enumerates important guidelines such as using clear and concise language, a consistent frame of reference, to speak in the present tense, and to limit descriptions to two or three sentences. See \autoref{fig:ai_describe_diagram} for AI Describer's system diagram and  \autoref{fig:ai_describer_examples} for example scenes and output. In addition to \keys{Alt} + \keys{D}, the user can tab to and select the ``Generate Description'' button, which uses a similar prompt but requests a structured JSON response from Gemini with mobility-related features, potential navigation obstacles for BLV pedestrians, safety risks, and a list of three follow-up questions that the user could ask in AI Chat, which are appended to the UI (\textit{e.g.,} \autoref{fig:ai_describer_examples}b). 


\subsubsection{AI Chat Agent}
\begin{figure*}[!htbp]
    \centering
    \includegraphics[width=0.5\textwidth]{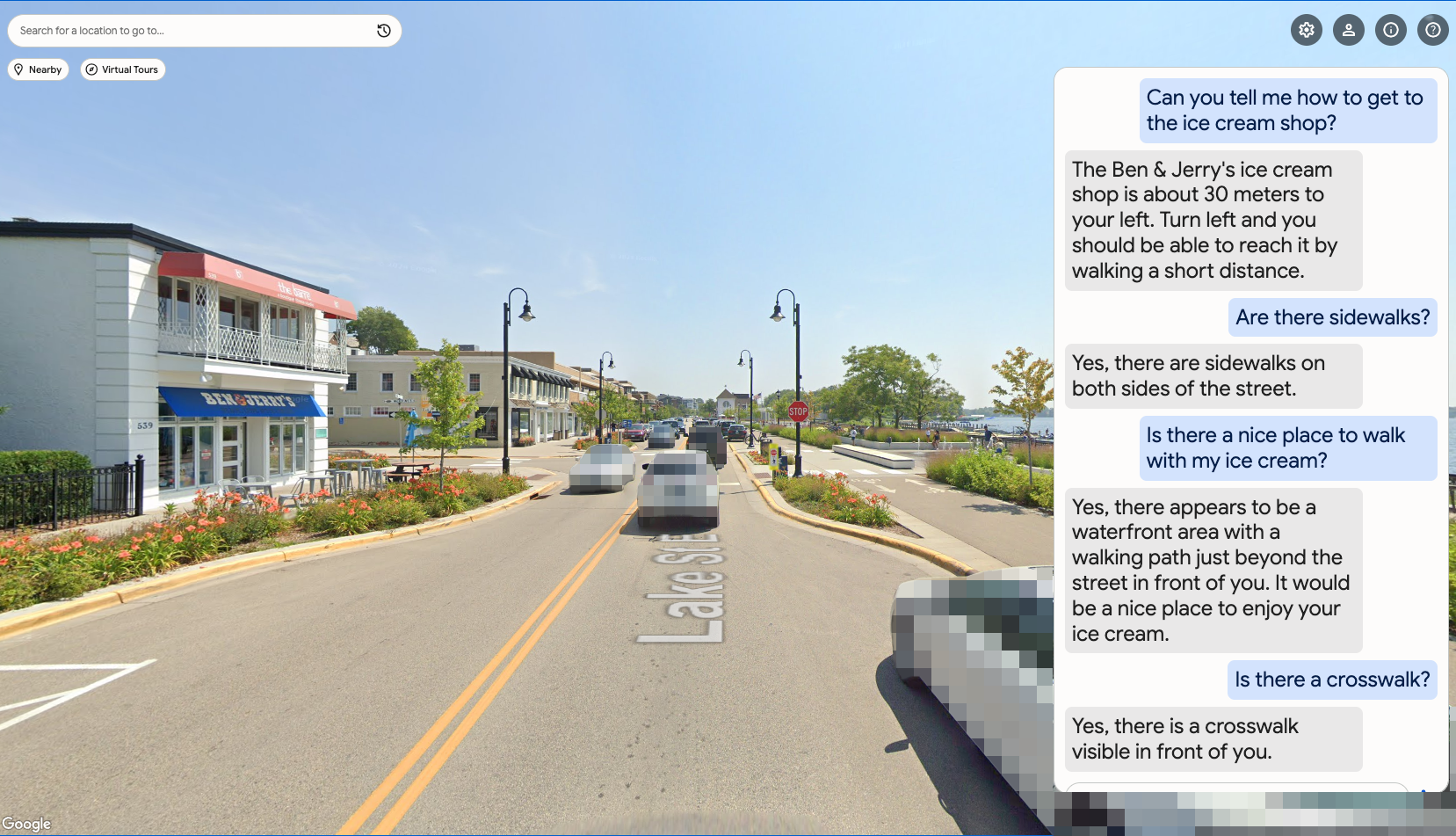}%
    \Description{A scene from the ice cream navigation task showing a two-lane street with a double yellow line and Ben and Jerry's on the left. The user is conversing with the AI Chat: "Can you tell me how to get to the ice cream shop?" AI response: "The Ben & Jerry's ice cream shop is about 30 meters to your left. Turn left and you should be able to reach it by walking a short distance." User: "Are there sidewalks?" AI: "Yes, there are sidewalks on both sides of the street." User: "Is there a nice place to walk with my ice cream?" AI: "Yes, there appears to be a waterfront area with a walking path just beyond the street in front of you. It would be a nice place to enjoy your ice cream." User: "Is there a crosswalk?" AI: "Yes, there is a crosswalk visible in front of you."}%
    \includegraphics[width=0.5\textwidth]{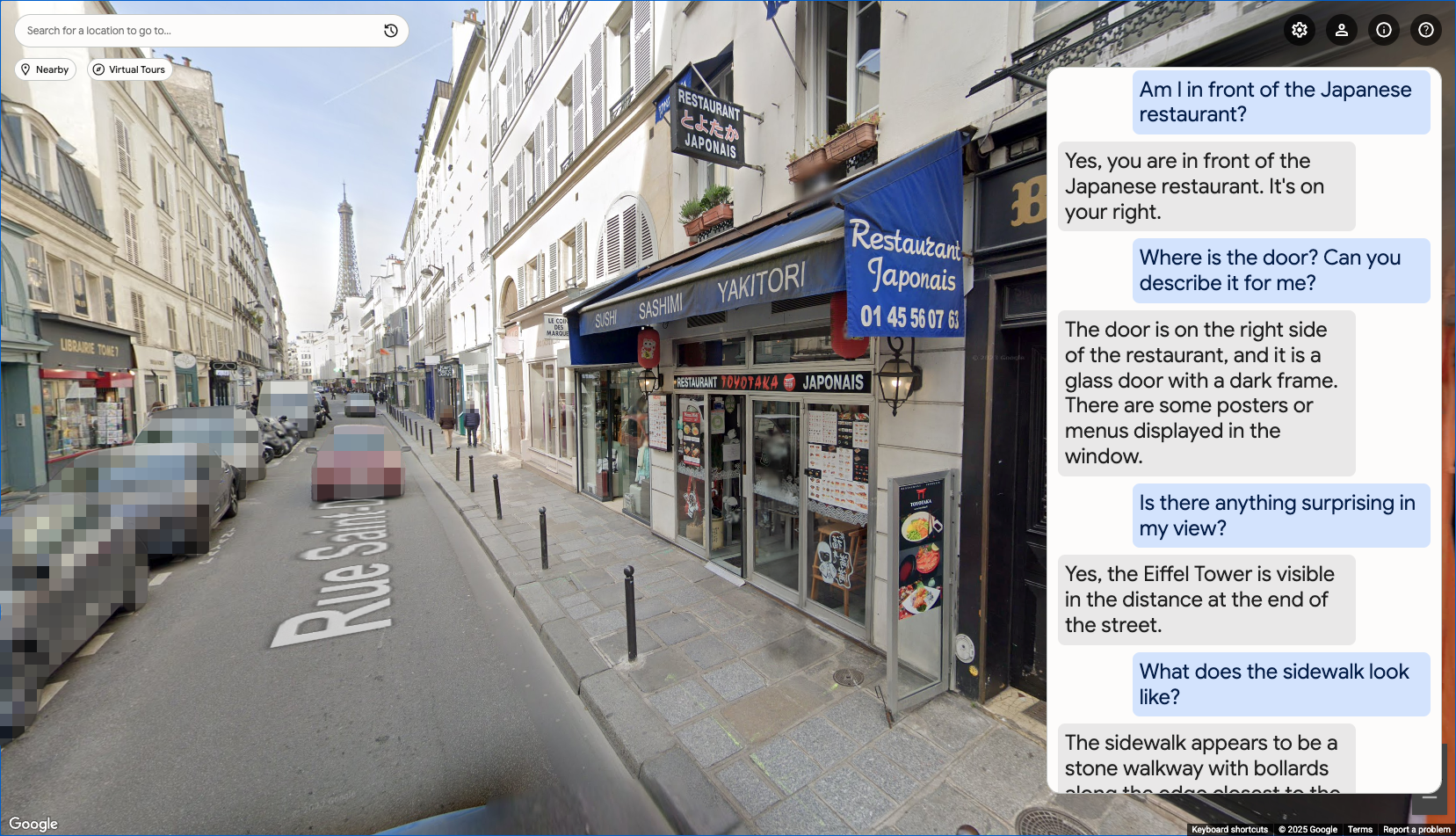}%
    \Description{A scene from the Japanese restaurant task showing a Parisian street, the Japanese restaurant on the right (with a blue awning), and the Eiffel tower in the distance. Here is the AI Chat shown on the right side of the screenshot: "Am I in front of the Japanese restaurant?" AI: "Yes, you are in front of the Japanese restaurant. It's on your right." User: "Where is the door? Can you describe it to me?" AI: "The door is on the right side of the restaurant, and it is a glass door with a dark frame. There are some posters or menus displayed in the window." User: "Is there anything surprising in my view?" AI: "Yes, the Eiffel Tower is visible in the distance at the end of the street." User: "What does the sidewalk look like?" AI: "The sidewalk appears to be a stone walkway with bollards along the edge closest to the street."}%
    \vspace{-1.0em}
    \caption{\textit{AI Chat} examples drawn from the open-world navigation tasks in Part 3 of our user study. See also the video demo.}
    \label{fig:ai_chat_examples}
\end{figure*}

Complementing AI Describer, the \textit{AI Chat Agent} allows for conversational interactions about the user's current and past views as well as nearby geography (Figures \ref{fig:ai_chat_diagram} and \ref{fig:ai_chat_examples}). The agent uses Google's \textit{Multimodal Live API}~\cite{google_multimodal_live_api}, which supports real-time interaction, function calling, and, crucially, retains memory of all interactions within a single session. When the user initiates a chat either via typing (\keys{Alt} + \keys{C}), speaking (\keys{Alt} + \keys{Spacebar}), or both, we begin a WebSocket connection that persists throughout the chat. Every pan and movement interaction is then sent to the AI Chat Agent along with a corresponding image of the user's current view and geographic context (\textit{e.g.,} nearby places, user's current heading, \textit{etc.}). The context window is set to a maximum of 1,048,576 input tokens, which is roughly equivalent to over 4k input images or ~8 average-length English novels. 

\begin{figure*}[!htbp]
    \centering
    \includegraphics[width=0.5\textwidth]{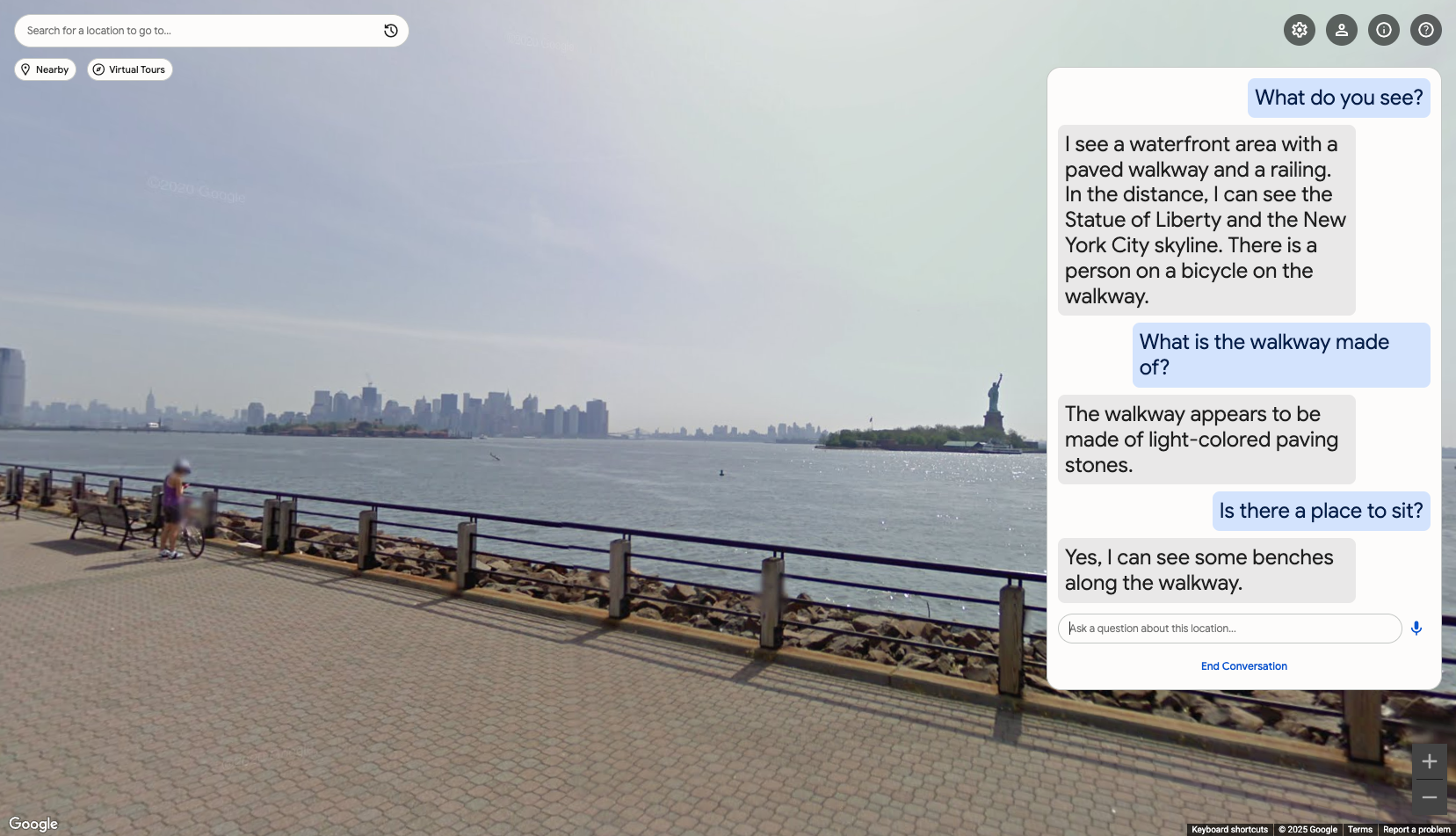}%
    \Description{A GSV screenshot showing the New York skyline, water, and the Statue of Liberty in the distance. There is an AI Chat panel on the right side, it reads: User: "What do you see?" AI: "I see a waterfront area with a paved walkway and a railing. In the distance, I can see the Statue of Liberty and the New York City skyline. There is a person on a bicycle on the walkway." User: "What is the walkway made of?" AI: "the walkway appears to be made of light-colored paving stones." User: "Is there a place to sit?" AI: "Yes, I can see some benches along the walkway."}%
    \includegraphics[width=0.5\textwidth]{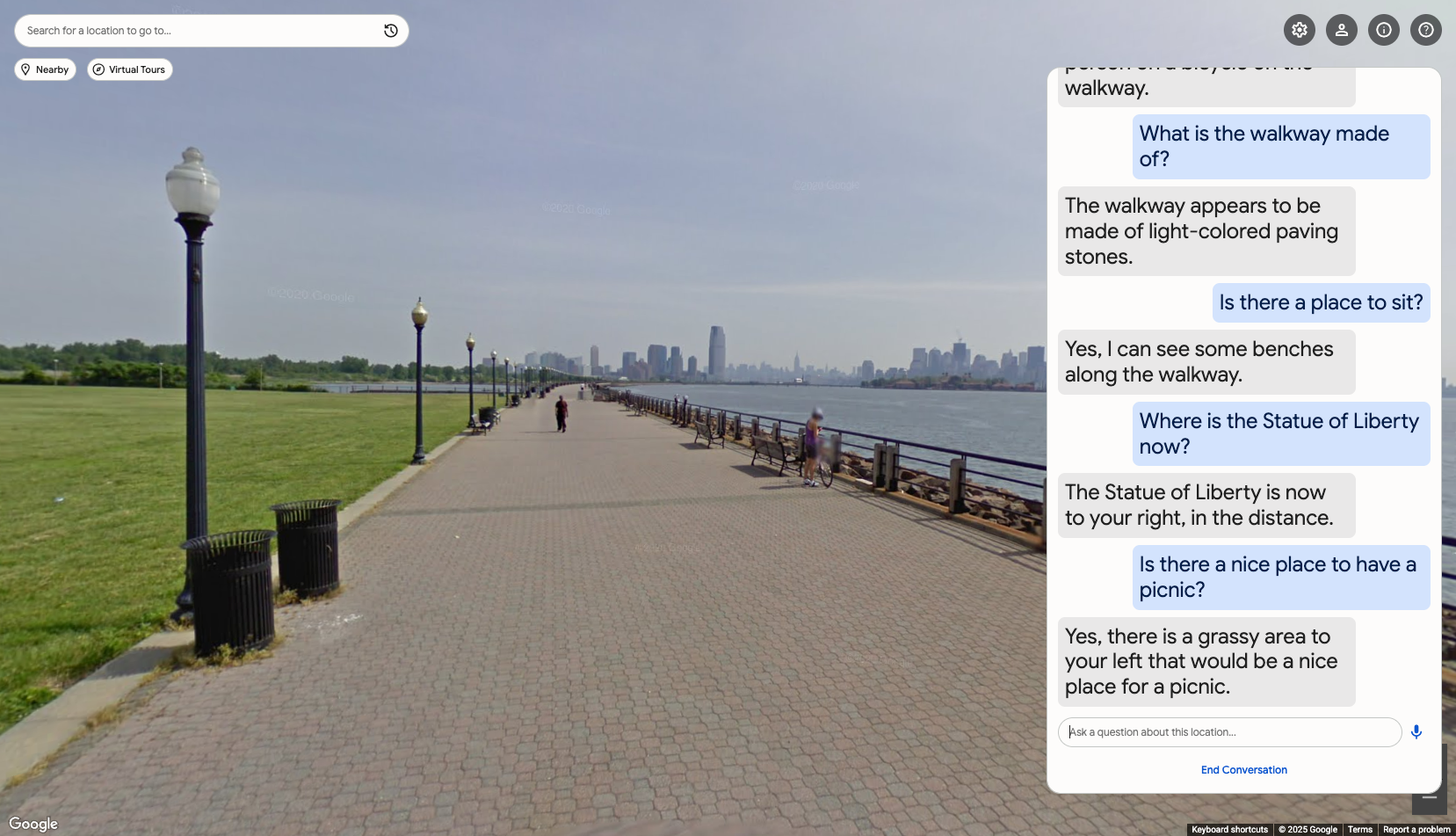}%
    \Description{Here, the user rotated their view with the keyboard, turning left. Now the user is facing down toward the path and a green grassy area is visible. The AI chat continues. User: "Where is the Statue of Liberty now?" AI: "The Statue of Liberty is now to your right, in the distance." User: "Is there a nice place to have a picnic?" AI: "Yes, there is a grassy area to your left that would be a nice place for a picnic."}%
    \vspace{-1.0em}
    \caption{Demonstrating spatial orientation support in \textit{AI Chat}. The user turned to the left in the second image.}
    \label{fig:ai_chat_spatial_orientation}
\end{figure*}

Because of this multimodal context window, the user can ask contextualized questions such as \myquote{Where is the bus stop?} and, if the bus stop existed in previous views or in the geographic data, the agent can appropriately respond with a contextualized answer \myquote{The bus stop is behind you, approximately 12 meters away.}---see \autoref{fig:ai_chat_spatial_orientation}. \revised{While the Live API retains conversational, navigational, and view history, we did not specifically tune nor supply an explicit spatial orientation model. Though we observed frequently correct spatial interpretations in our user study, more work is necessary to investigate how to achieve robust spatial orientation in MLLMs, especially with dynamic movement and longer, more complex navigational histories (see \ref{sec:disc_orientation} in the Discussion).}

All AI Chat responses are spoken either natively by \sysname using ``text-to-speech'' (if native voicing mode is turned on) or the user's screen reader via ARIA live messages. While both typing and speaking results in the same underlying function calls, in practice, we found that using speech input enabled a more seamless, multimodal input experience where the user could pan or move and immediately vocalize a question such as \myquote{What am I looking at now?} or \myquote{Is there anything interesting here?}. We use two audio earcons to signify when the mic is turned on/off.

\textbf{Commands.} In addition to asking questions, the user can also control \sysname via commands spoken or typed to the agent, including to pan, move, and jump (\textit{e.g.,} ``turn left'' or ``move forward''; see \autoref{tab:ai_chat_command_functions}). This is implemented via the Live API's \texttt{tools} method, which takes function declarations and descriptions as input. We parse the agent's response and call the appropriate functions. Even with these simple controls, we found that natural language commands were possible: instead of \myquote{Turn me left by 45 degrees,} the user could say \myquote{Turn left a little bit.} Even further and more surprisingly, more complex command behavior emerged where the model incorporated context to enable semantic commands like \myquote{Turn me towards the cafe.} See video demo and \autoref{fig:ai_chat_diagram}c.


\begin{table}[h!]
\centering
\footnotesize
\renewcommand{\arraystretch}{1.2}
\begin{tabularx}{\columnwidth}{@{} l X @{}}
\toprule
\textbf{Function Name} & \textbf{Function Description} \\
\midrule
\rowcolor[HTML]{EFF2F6}
\texttt{moveBackward} & Move the user backward opposite the current heading to the previous Google Street View panorama \\
\texttt{moveForward} & Move the user forward along the current heading to the next Google Street View panorama \\
\rowcolor[HTML]{EFF2F6}
\texttt{moveToIntersection} & Move the user's location to the next intersection \\
\texttt{turnLeft45} & Turn the user's heading a bit to the left (-45 degrees) \\
\rowcolor[HTML]{EFF2F6}
\texttt{turnLeft90} & Turn the user's heading to the left (-90 degrees) \\
\texttt{turnRight45} & Turn the user's heading a bit to the right (45 degrees) \\
\rowcolor[HTML]{EFF2F6}
\texttt{turnRight90} & Turn the user's heading to the right (90 degrees) \\
\texttt{turnAround} & Turn the user's heading around (180 degrees) \\
\bottomrule
\end{tabularx}
\caption{Function declarations and descriptions provided to the \textit{AI Chat Agent} for controlling \sysname via chat.}
\label{tab:ai_chat_command_functions}
\end{table}


\subsubsection{AI Tour Guide}
\begin{figure*}[!htbp]
    \centering
    \includegraphics[width=0.333\textwidth]{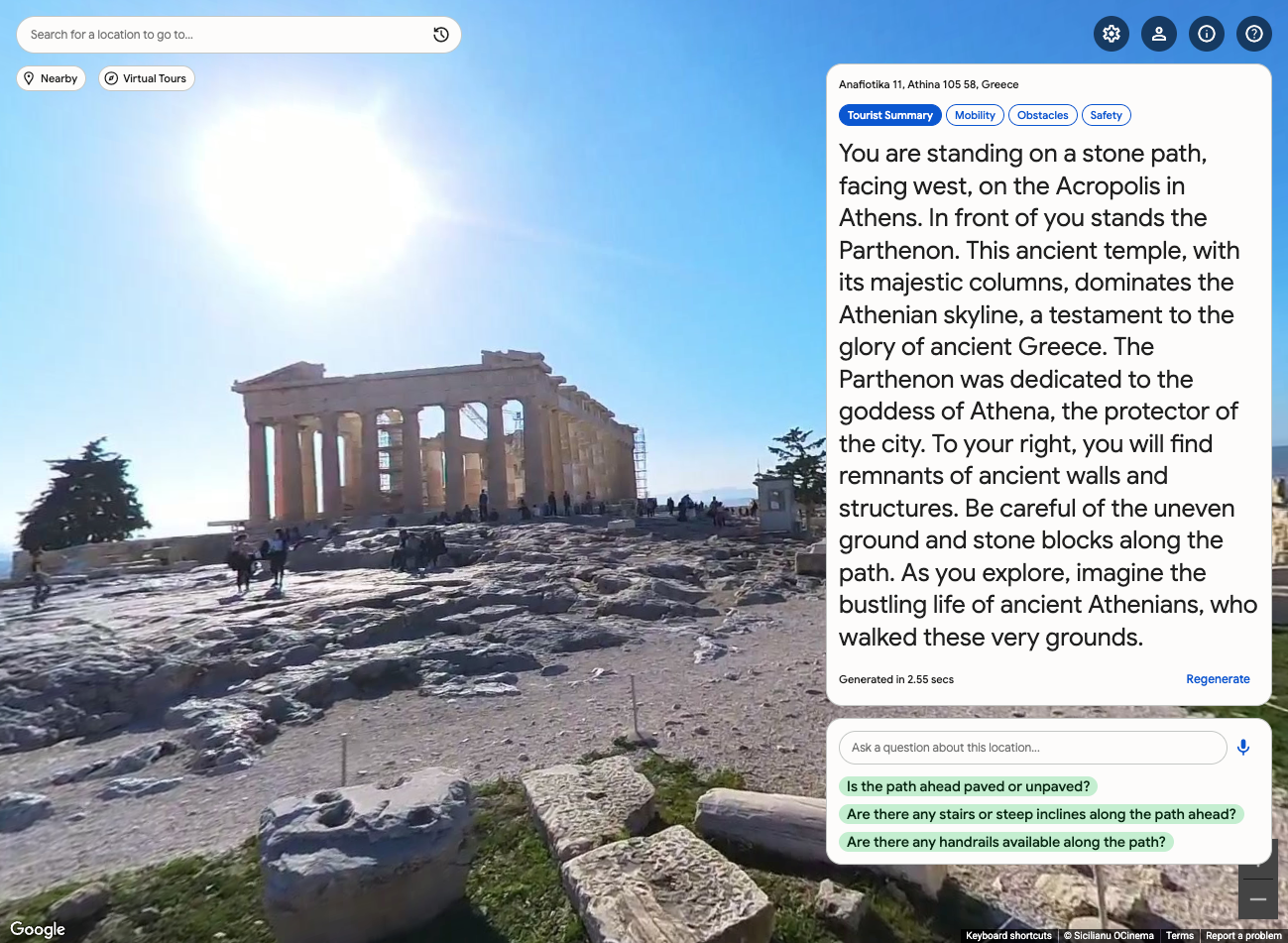}%
    \Description{AI tourist-generated description: "You are standing on a stone path, facing west, on the Acropolis in Athens. In front of you stands the Parthenon. This ancient temple, with its majestic columns, dominates the Athenian skyline, a testament to the glory of ancient Greece. The Parthenon was dedicated to the goddess of Athena, the protector of the city. To your right, you will find remnants of ancient walls and structures. Be careful of the uneven ground and stone blocks along the path. As you explore, imagine the bustling life of ancient Athenians, who walked these very grounds."}%
    \includegraphics[width=0.333\textwidth]{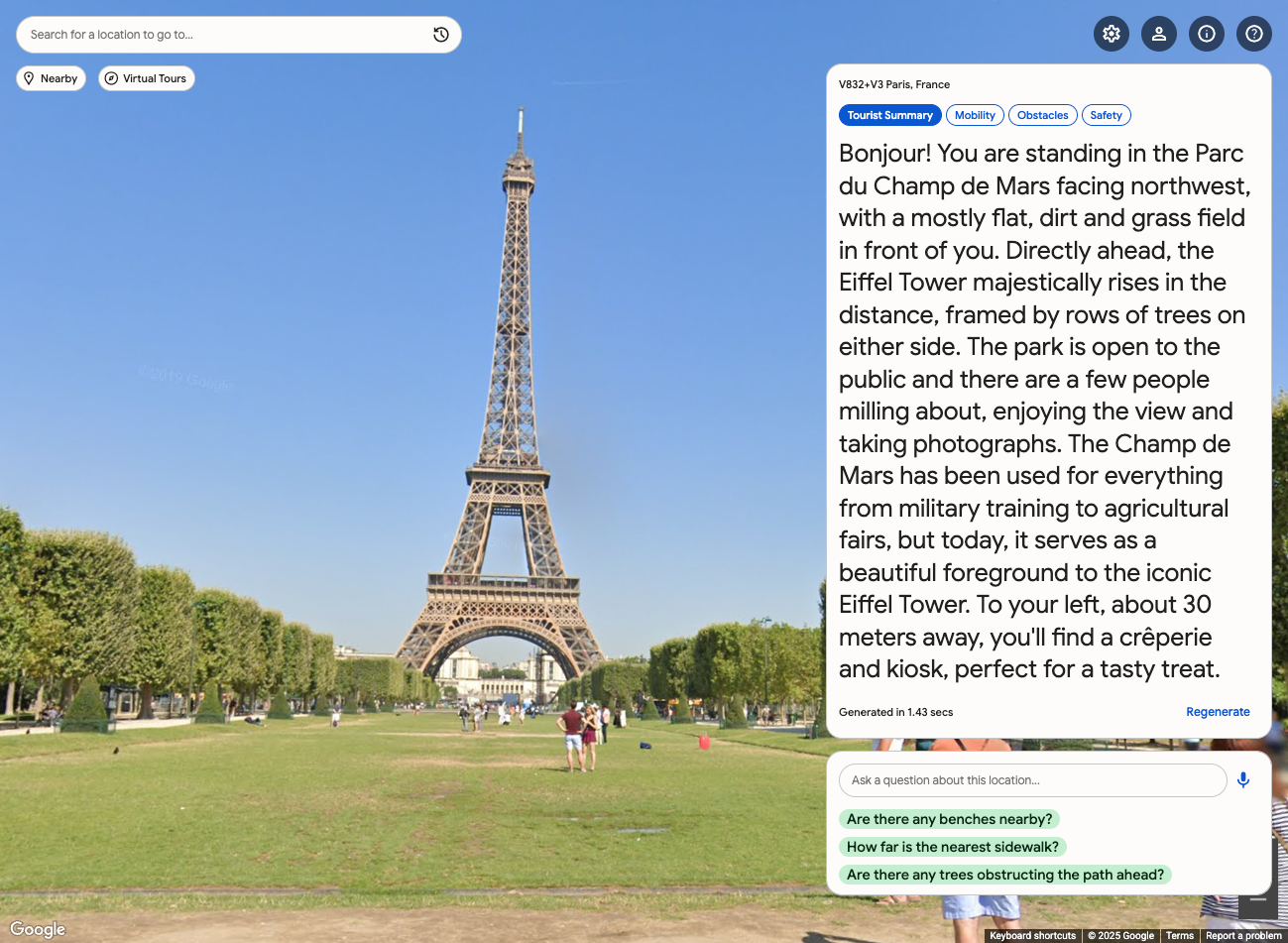}%
    \Description{AI tourist-generated description: "Bonjour! You are standing in the Parc du Champ de Mars facing northwest, with a mostly flat, dirt and grass field in front of you. Directly ahead, the Eiffel Tower majestically rises in the distance, framed by rows of trees on either side. The park is open to the public and there are a few people milling about, enjoying the view and taking photographs. The Champ de Mars has been used for everything from military training to agricultural fairs, but today, it serves as a beautiful foreground to the iconic Eiffel Tower. To your left, about 30 meters away, you'll find a crêperie and kiosk, perfect for a tasty treat."}%
    \includegraphics[width=0.333\textwidth]{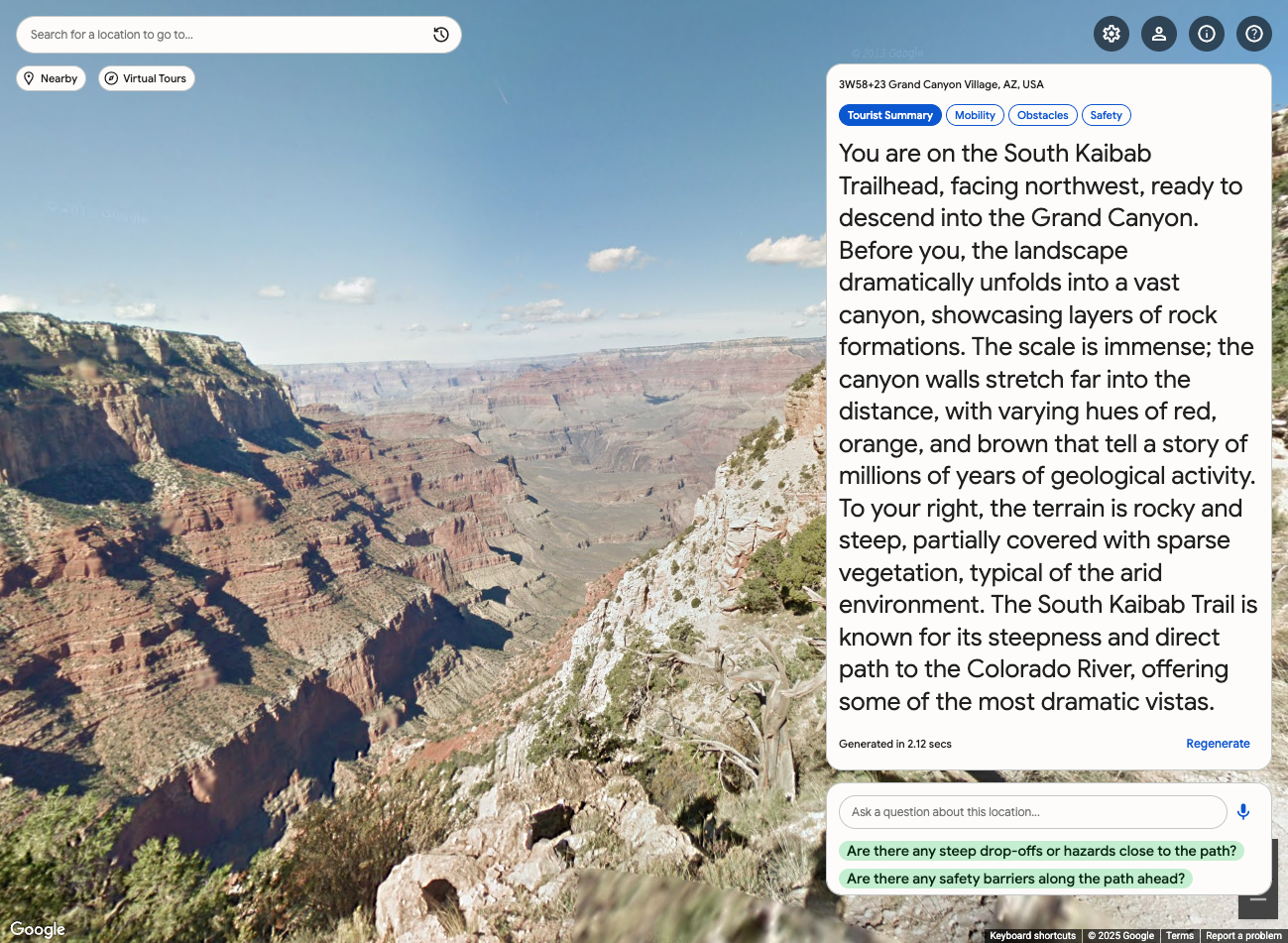}%
    \Description{AI tourist-generated description of the Grand Canyon: "You are on the South Kaibab Trailhead, facing northwest, ready to descend into the Grand Canyon. Before you, the landscape dramatically unfolds into a vast canyon, showcasing layers of rock formations. The scale is immense; the canyon walls stretch far into the distance, with varying hues of red, orange, and brown that tell a story of millions of years of geological activity. To your right, the terrain is rocky and steep, partially covered with sparse vegetation, typical of the arid environment. The South Kaibab Trail is known for its steepness and direct path to the Colorado River, offering some of the most dramatic vistas."}%
    \vspace{-0.5em}
    \caption{Virtually visiting the Parthenon, the Eiffel Tower, and the Grand Canyon using the \textit{AI Tour Guide}. With \sysname, users can virtually travel to over 220 billion street view images spanning 100 countries \cite{Google_StreetView_Stats}.}
    \label{fig:tour_guide_examples}
\end{figure*}

Finally, the \textit{AI Tour Guide} subsystem is essentially AI Describer but with a specialized prompt directing the Gemini model to act as an \myquote{expert tour guide for blind or low-vision virtual tourists} (\autoref{fig:ai_describe_diagram}b). The prompt instructs the AI to combine clear visual descriptions with tourist-focused content, including historical facts, cultural significance, architectural styles, interesting anecdotes, nearby popular attractions, and descriptions of human activity in the scene to create a more engaging and informative experience, akin to a real guided tour (\textit{e.g.,} \autoref{fig:tour_guide_examples}).

\subsection{Other Controls}
Beyond the commands already mentioned, the user can press \keys{Alt} + \keys{W} to hear their current address and heading, \keys{Alt} + \keys{I} for nearby intersections, \keys{Alt} + \keys{N} for nearby places, and \keys{Alt} + \keys{P} to hear the date and photographer of the current pano (\textit{e.g.,} \myquote{This Street View image was taken on February 2025 by Google}). See \autoref{tab:hotkeys}.

\section{Implementation}

\sysname's frontend is implemented in TypeScript, Sass (SCSS), and HTML with the open-source libraries \textit{Lit} for declarative UI components~\cite{lit}, \textit{RxJS} for reactive state management~\cite{rxjs}, and \textit{MobX} for efficient state observation and synchronization~\cite{mobx}. The backend is implemented in a custom lightweight \textit{Python} server (similar to \textit{Flask}), which serves static app resources. No data is currently stored on our web server---key functionality, including AI implementations, are done via client-side implementations and external API calls, all which use the Google Cloud infrastructure.

For nearby places, road information, and the interactive streetscape images themselves, we use the \textit{Google Maps Places}~\cite{google_places_api}, \textit{Roads}~\cite{google_roads_api}, and \textit{StreetView}~\cite{google_streetview_api} APIs, respectively. \revised{Data used by permission from Google.} \sysname is designed to be natively voiced---toggable to work with or without screen readers \revised{with a configurable speech output rate}---which is implemented via Google Cloud's ``\textit{text-to-speech}'' API~\cite{google_tts_api}. The voice recognition and transcription is implemented with Google's ``\textit{speech-to-text}'' API~\cite{google_stt_api}. The AI subsystems were implemented with Google's \textit{Vertex AI} platform and \textit{Gemini 2.0 Flash Model} \cite{google_vertex_ai}. To more easily discern status update messages \textit{vs.} AI Chat responses, status messages are read with a male voice (``en-US-Wavenet-J'') and the chat with a female voice (``en-US-Wavenet-G'')---see \cite{google_tts_languages}. For efficiency, all API calls are cached in the browser using a custom, location-based \textit{IndexedDB} instance.
\section{User Study}
To evaluate \sysname and explore the potential of an accessible, AI-driven street view experience, we conducted an in-person lab study with eleven blind participants. \revised{The study tasks are informed by our co-design experiences as well as relevant prior work in BLV navigation, including our own (\textit{e.g.,} SoundScape\cite{Microsoft_Soundscape_2023}). We focus on POI investigations and virtual route exploration.}


\begin{table*}[h!]
\centering
\footnotesize
\renewcommand{\arraystretch}{0.8}
\setlength{\tabcolsep}{3pt}
\setlength{\extrarowheight}{0pt}   
\renewcommand{\tabularxcolumn}[1]{>{\raggedright\arraybackslash}p{#1}}
\begin{tabularx}{\textwidth}{@{} >{\centering\arraybackslash}p{5cm} >{\raggedright\arraybackslash}X >{\raggedright\arraybackslash}X @{}}
\toprule
\multicolumn{1}{@{}c@{}}{\textbf{StreetView}} & \textbf{Task} & \textbf{Questions} \\
\midrule
\rowcolor[HTML]{EFF2F6}
\raisebox{-\totalheight}{%
    \includegraphics[width=5cm,keepaspectratio]{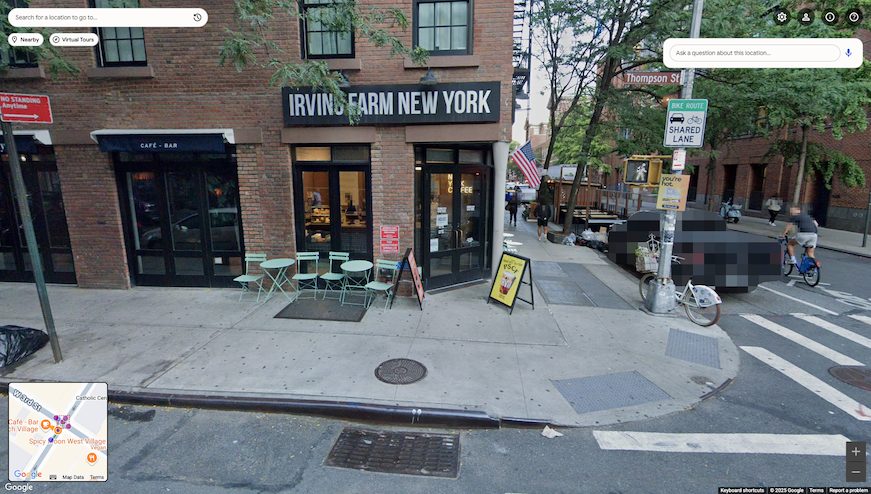}
    \Description{The first POI image, which was part of the tutorial, showing a cafe called Irving Farm New York in front of the user at the corner of a brick building. There are green tables and chairs outside. There are two curb ramps. Here's an example AI-generated description from StreetViewAI: "You are standing on a crosswalk facing a brick building on the corner of the block. To your left, set back from the corner, is a business called "Irving Farm New York," its name displayed on a sign above the entrance. There are double doors partially blocked by an advertisement sign on the sidewalk. To the right, Thompson Street continues towards the Northwest, with a parked black pickup truck and a bicyclist visible further down the block."}
}
& 
\vspace{0pt}\textbf{Tutorial task: Coffee Shop}

We will virtually teleport to a coffee shop in New York called Irving Farm New York. Your goal is to evaluate the coffee shop and the surroundings to help you navigate there in real life. 
& 
\vspace{0pt}\begin{itemize}[noitemsep,topsep=0pt,parsep=0pt,partopsep=0pt,leftmargin=*]
\item Are there chairs to sit on outside the café?
\item Are there obstacles on the sidewalk?
\item What road are you on?
\item Are we at an intersection? If so, what?
\item What else is nearby me?
\item When was the photograph taken?
\item <Ask your own question>
\end{itemize}
\\
\raisebox{-\totalheight}{%
  \includegraphics[width=5cm,keepaspectratio]{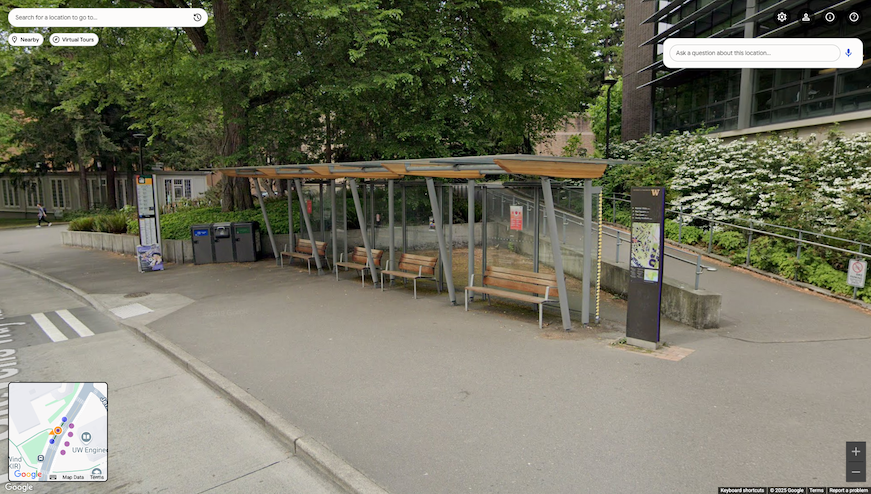}
  \Description{The bus stop POI investigation with a sheltered bus stop set back from the sidewalk. Here's an example AI-generated description from StreetViewAI: "The scene appears to be on a sidewalk near a university campus. Directly in front of you is a bus stop with several benches under a metal awning. To the right of the bus stop, a paved walkway leads up to a modern building with large windows."}
}
& 
\vspace{0pt}\textbf{Bus Stop}

You are visiting a new neighborhood and need to learn about a bus stop. Use StreetViewSR to interact with the scene and learn about the bus stop features that would help you travel there in real life (e.g., navigational features, landmarks, etc.).
&
\vspace{0pt}\begin{itemize}[noitemsep,topsep=0pt,parsep=0pt,partopsep=0pt,leftmargin=*]
\item What does the bus stop look like?
\item Are there benches and a bus shelter?
\item What are some surrounding navigational landmarks?
\item Are there garbage cans?
\item What are some nearby buildings?
\end{itemize}
\\
\rowcolor[HTML]{EFF2F6}
\raisebox{-\totalheight}{%
  \includegraphics[width=5cm,keepaspectratio]{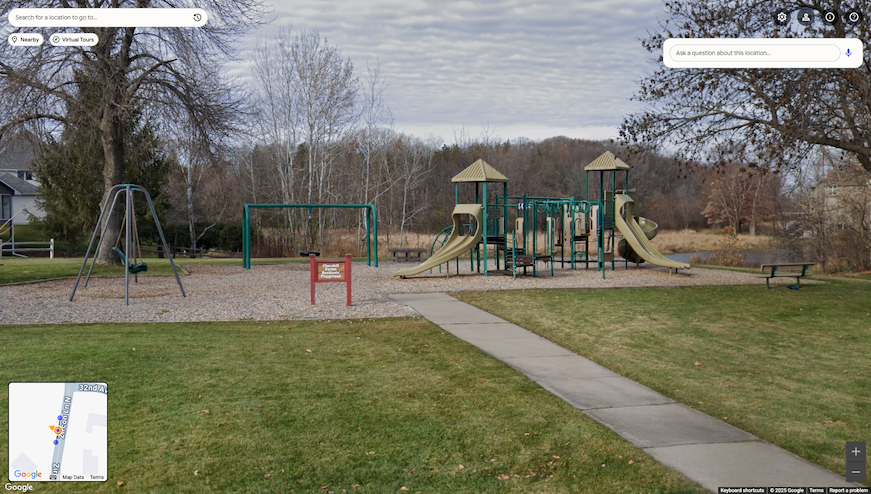}
  \Description{The third POI investigation: image shows a playground in the distance with a path flanked by grass. An example AI Describe description from StreetViewAI: "You are standing on a sidewalk, looking across a grassy lawn toward a playground. A concrete path cuts across the lawn from your right to the playground equipment, which includes swings and slides. In the distance, behind the playground, is a grove of trees."}
}
& 
\vspace{0pt}\textbf{Playground}

You are visiting your sister's family and want to bring your two young nieces to a new playground. Your goal is to learn about the playground before going. What kind of equipment does it have? Does it seem safe?

&
\vspace{0pt}\begin{itemize}[noitemsep,topsep=0pt,parsep=0pt,partopsep=0pt,leftmargin=*]
\item What does the playground look like?
\item Are there slides and/or swings?
\item What kind of ground cover is there?
\item Are there benches to sit on?
\item Are there sidewalks?
\item Does the street look busy?
\item What kind of neighborhood does it look like?
\item What kind of houses are nearby?
\end{itemize}
\\
\raisebox{-\totalheight}{%
  \includegraphics[width=5cm,keepaspectratio]{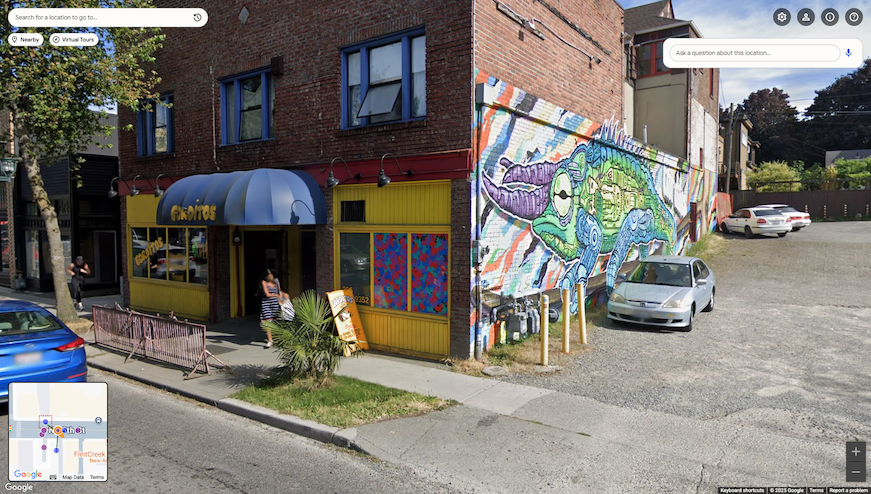}
  \Description{The last POI investigation. The image shows a Mexican restaurant with a blue awning and mural. There is a parking lot on the right side and a bike rack outside the entrance. Here's an example StreetViewAI AI Describe description: "The scene is a street view in the Greenwood neighborhood of Seattle. Directly in front of you is a two-story brick building with a business on the ground floor that has yellow trim, a mural, and a blue awning that appears to say "Gorditos." To the left of the building, a blue car is parked along the curb."}
}
& 
\vspace{0pt}\textbf{Mexican Restaurant}

You are meeting a friend at a Mexican restaurant unfamiliar to both of you. Your friend heard there was something interesting about the building and also asked about whether to drive or bike (\textit{e.g.,} is there a parking lot nearby or a bike rack for their bike?).
&
\vspace{0pt}\begin{itemize}[noitemsep,topsep=0pt,parsep=0pt,partopsep=0pt,leftmargin=*]
\item What does the building look like?
\item What is unique about the building?
\item Is there a parking lot nearby?
\item Is there a bike rack?
\item Is there a sidewalk in front of the restaurant?
\end{itemize}
\\
\bottomrule
\end{tabularx}
\caption{The Part 2 ``POI Investigation'' tasks and questions for participants. For each scenario, we invited participants to explore the location on their own and ask their own questions before providing our specific list of questions. If the participant did not immediately know the answer to a posed question, they were encouraged to use \sysname to find out.}
\label{tab:poi_tasks_table}
\end{table*}

\subsection{Participants}
We recruited eleven Participants (6 men, 5 women) using mailing lists to local blind and low-vision organizations, newsletter posts, and a contact list from previous studies. Participants ranged in age from 20--66+. All used white canes for mobility; two also used guide dogs. For technology, all participants used screen readers, predominantly \textit{JAWS}~\cite{JAWS_ScreenReader} on \textit{Windows} and \textit{VoiceOver} on \textit{iOS}. Most reported experience with at least one mapping tool, most commonly \textit{Google} or \textit{Apple Maps} but also blind-specific navigation tools such as \textit{BlindSquare}~\cite{BlindSquare}, \textit{VoiceVista}~\cite{VoiceVista}, and \textit{GoodMaps}~\cite{GoodMaps}. While technology experience and comfort with laptops (and typing) varied, seven participants had medium-to-high familiarity with AI tools, most commonly \textit{Be My AI}~\cite{Be_My_AI}, \textit{ChatGPT}, and \textit{Seeing AI}~\cite{Seeing_AI}. Three wore Meta AI glasses but indicated using the camera+AI features rarely and, instead, relying on the convenient headphone output for their screen readers. Though a few participants had heard of GSV, none reported experience with any streetscape tools, highlighting an opportunity for this population. 

\begin{table*}[h!]
\centering
\footnotesize
\renewcommand{\arraystretch}{1.2} 
\setlength{\tabcolsep}{3pt}
\renewcommand{\tabularxcolumn}[1]{>{\raggedright\arraybackslash}p{#1}}
\begin{tabularx}{\textwidth}{@{} l >{\raggedright\arraybackslash}p{0.6cm} >{\raggedright\arraybackslash}p{0.6cm} >{\raggedright\arraybackslash}p{1.5cm} >{\raggedright\arraybackslash}p{1.2cm} >{\raggedright\arraybackslash}X >{\raggedright\arraybackslash}X >{\raggedright\arraybackslash}p{1cm} >{\raggedright\arraybackslash}X >{\raggedright\arraybackslash}p{1.5cm} @{}}
\toprule
\textbf{PID} & \textbf{Gen.} & \textbf{Age} & \textbf{Vision} & \textbf{Mobility Aid} & \textbf{Screenreader} & \textbf{Mapping Tools} & \textbf{AI Fam.} & \textbf{AI Tools} & \textbf{Streetscape Tools} \\
\midrule
\rowcolor[HTML]{EFF2F6}
1 & M & 18-25 & Blind & White cane & JAWS/NVDA on Windows; VoiceOver on iOS & BlindSqaure, GoodMaps, Google Maps, Lazarillo, Voice Vista & High & BeMyAI, ChatGPT, Copilot, Gemini, Image Describers & Never used \\
2 & W & 56-65 & Low vision & White cane & JAWS/NVDA; VoiceOver on iOS; Braille display & BlindSquare, Google and Apple Maps & High & ChatGPT, Meta Glasses, VoiceOver image description & Never used \\
\rowcolor[HTML]{EFF2F6}
3 & M & 18-25 & Blind, some light detection & White cane & JAWS/NVDA; VoiceOver on iOS; TalkBack on Android & BlindSquare, Google Maps & High & Gemini, TalkBack image description & Never used \\
4 & W & 46-55 & Low vision & White cane & JAWS; VoiceOver on iOS & Google/Apple Maps & Low & None & Never used \\
\rowcolor[HTML]{EFF2F6}
5 & M & 26-35 & Blind & White cane & JAWS/NVDA, VoiceOver on iOS, Braille display & Google/Apple Maps, VoiceVista & High & Be My AI, Copilot, Seeing AI, VoiceOver image description & Never used \\
6 & W & 46-55 & Blind & White cane & JAWS; VoiceOver on iOS & Google Maps (but rarely) & Low & None & Never used \\
\rowcolor[HTML]{EFF2F6}
7 & W & 66+ & Blind, some light detection & White cane & JAWS/Fusion; VoiceOver on iOS & Google/Apple Maps & Low & Tried Be My AI in class & Never used \\
8 & W & 26-35 & Blind, some light detection & White cane & JAWS; VoiceOver on iOS & BlindSquare, Google Maps with Voice Guidance & High & Be My AI, Seeing AI & Never used \\
\rowcolor[HTML]{EFF2F6}
9 & M & 26-35 & Blind & White cane & JAWS; VoiceOver on iOS & Google Maps with Voice Guidance & High & AIRA (but not with AI), Be My AI, ChatGPT & Never used \\
10 & M & 66+ & Blind, some light detection & White cane, guide dog & JAWS/Fusion; VoiceOver on iOS & BlindSquare, GoodMaps, VoiceVista & Medium & ChatGPT, Meta Glasses, Perplexity, Seeing AI & Never used \\
\rowcolor[HTML]{EFF2F6}
11 & M & 66+ & Blind & White cane, guide dog & JAWS; VoiceOver on Mac and iOS & BlindSquare, GoodMaps, Google Maps, VoiceVista & Medium & AIRA (but not with AI), Be My AI, Copilot, Meta Glasses & Never used \\
\bottomrule
\end{tabularx}
\caption{Participant demographics, mobility aids, and technology experience. \textit{AI Fam.} = AI Familiarity.}
\label{tab:participant_table}
\end{table*}

\subsection{Procedure}

The single-session, in-person lab study consisted of four parts and lasted 1.5-2 hrs: (1) a formative interview inquiring about the use of screen readers, navigation tools, maps, and AI; (2) a brief introduction and tutorial task with \sysname followed by points-of-interest (POI) investigations; (3) two open-world navigation tasks in pre-selected areas followed by a third open-world navigation at a location of the participant's choice; (4) and a debrief interview. For our selected geographies, we specifically chose locations that would be unfamiliar to the participant so they would rely on \sysname rather than their own memories. See the supplementary materials for the full protocol.

Because of the complexity and open-endedness of the study tasks, the varying technical proficiency of the participants, and because we wanted to observe users naturally engage with \sysname and ask questions of the AI, not all participants made it to all tasks. The protocol was intentionally designed to accommodate different task completion speeds, and we skipped the last navigation task (as necessary). Prior to the study, participants filled out a demographic form asking about vision, screen reader usage, and assistive technology.

\textbf{Part 1: Formative Interview.} In Part 1, we asked participants about their use of screen readers and mapping tools, how they navigate physical environments, plan routes, and use accessible navigation tools for support (\textit{e.g.,} BlindSquare) as well as their use of emerging AI tools (\textit{e.g.,} Be My AI, ChatGPT). Finally, we discussed their familiarity and use, if any, of streetscape imagery tools.

\textbf{Part 2: POI Investigations.} Part 2 focused on using \sysname to evaluate destinations (\textit{i.e.,} points-of-interest or POIs). Our research questions included: \textit{What} are BLV people interested in knowing about destinations? \textit{What} questions do they ask? \textit{How} well does our tool perform? We began with a brief introduction of what street view imagery is, how the data is collected, and a tutorial task where participants virtually visited a cafe in New York and learned basic controls, hot keys, and were asked to answer questions such as \myquote{Are there chairs to sit on outside the cafe?} and \myquote{Are there obstacles on the sidewalk?}. Participants were then invited to pan and move via the arrow keys and answer questions about the current intersection and other nearby places. We encouraged participants to ask their own questions and to be adventurous.

Following the tutorial, participants investigated three POIs with scenarios drawn from literature (\textit{e.g.,} \cite{Hara_BusStopLandmarks_TACCESS15, Hara_ImprovingPublicTransitAccess_ASSETS13, Szpiro_FindingAStore_UbiComp16, Kulkarni_BusStopCV_ASSETS23, Jiang_MakingUrbanArtAccessible_UrbanAccess24}) as well as our co-design sessions and, crucially, could only be addressed via visual inquiries with \sysname: (1) investigating features at a bus stop; (2) assessing a playground for fun and safety, and (3) examining an unfamiliar Mexican restaurant and surroundings. Following open exploration time, we asked participants about the scene. See \autoref{tab:poi_tasks_table}.



\begin{figure*}[h!]
\centering
\footnotesize
\setlength{\tabcolsep}{1pt}
\begin{tabular}{cccc} 
\textbf{West} & \textbf{North} & \textbf{East} & \textbf{South} \\ 
\includegraphics[width=0.25\textwidth,keepaspectratio]{figures/POITask2-Playground-3.png} 
\Description{Showing the west view at the playground POI task. Here's an example StreetViewAI Describe description: "}
&
\includegraphics[width=0.25\textwidth,keepaspectratio]{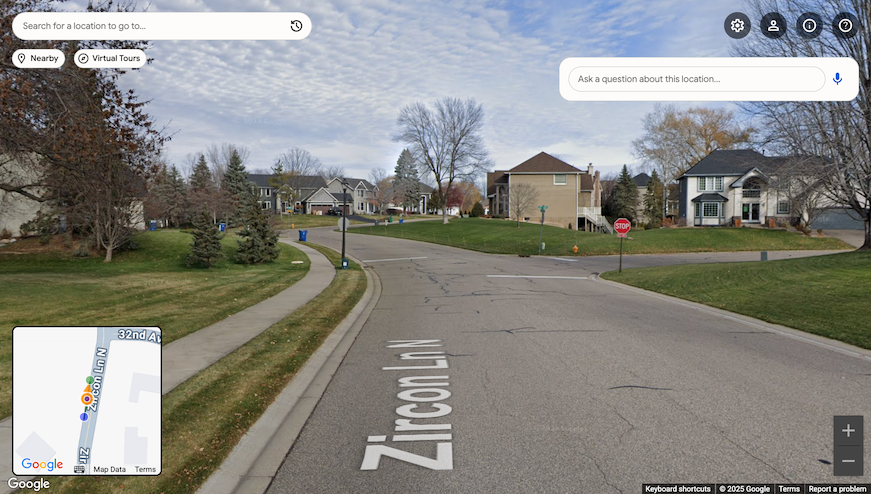}
\Description{Showing the north view at the playground POI task. Here's an example StreetViewAI AI Describe description: "You are on a residential street facing north, with a sidewalk to your left. About 50 feet in front of you, the road curves at an intersection with a stop sign on the other side. There are houses on both sides of the street."}
& 
\includegraphics[width=0.25\textwidth,keepaspectratio]{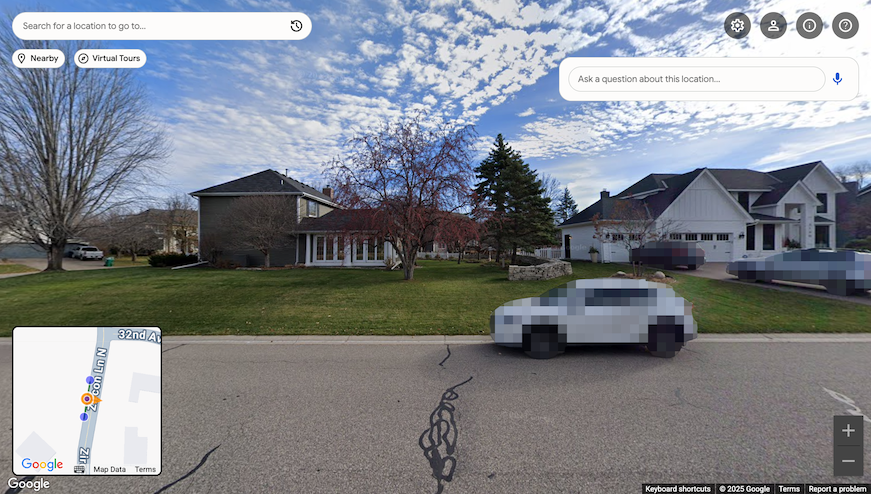}
\Description{Showing the east view at the playground POI task. Here's an example StreetViewAI AI Describe description: "You are on a residential street, facing east. To your right on the street is a white car, and beyond that is a house with a white facade and a dark roof. To your left is a house that appears to be darker. There does not appear to be a sidewalk."}
& 
\includegraphics[width=0.25\textwidth,keepaspectratio]{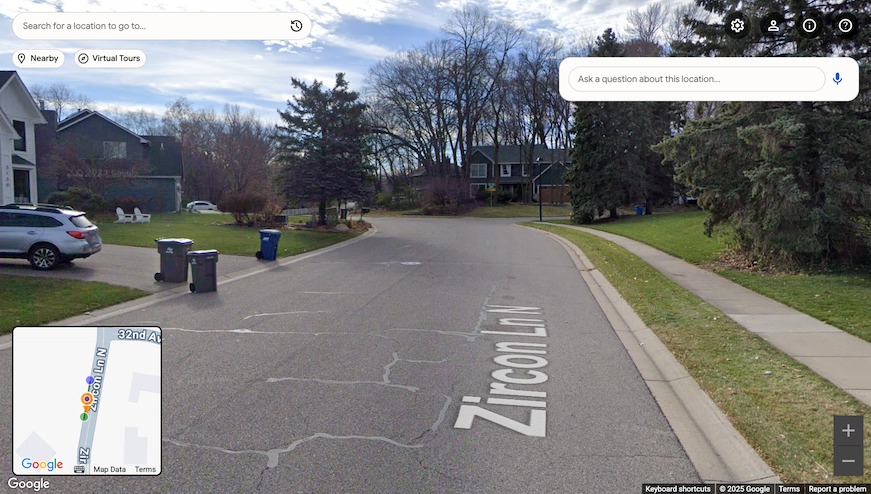}
\Description{Showing the south view at the playground POI task. Here's an example StreetViewAI AI Describe description: "You are on a residential street facing south. To the left, there's a large two-story house with a white facade and an attached garage, and a silver SUV is parked in its driveway. To your right, a sidewalk runs parallel to the road, bordered by a grassy area and tall evergreen trees."}
\\ 
\end{tabular}
\vspace{-1.0em}
\caption{\revised{Directional views of the Part 2 ``Playground Scenario''. Unlike the other POI study tasks, the playground task required participants to pan beyond the initial view to understand the surrounding context (\textit{e.g.,} residential neighborhood, quiet street).}}
\label{fig:poi_playground_views}
\end{figure*}

\textbf{Part 3: Open-world Navigation.} In Part 3, participants completed open-world navigation tasks to virtually explore potential walking routes in two geographies that we selected---an ice cream shop in Minneapolis and a Japanese restaurant in Paris---and then in an area familiar to the participant (\textit{e.g.,} their own neighborhood). \revised{This sort of pre-journey planning and virtual route rehearsal is a cornerstone of \textit{Orientation and Mobility} (O\&M) training for BLV individuals~\cite{Banduka_OandM_ASSETS21}.} Here, our research questions included: \textit{How} do BLV users combine multiple information sources to navigate? \textit{What} visual features in a scene are they most interested in? \textit{What} parts of \sysname are most useful to supporting open navigation?

\begin{figure*}[tb] 
  \centering 
  \includegraphics[width=\textwidth]{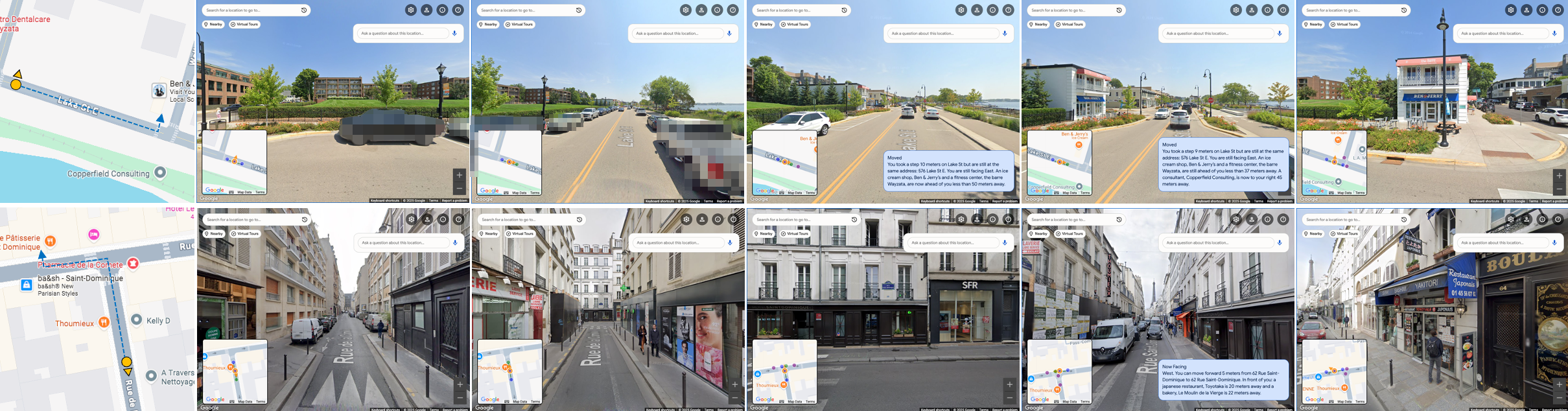}
  \vspace{-2.0em} 
  \caption{In Part 3, participants used \sysname to complete two open-world navigation tasks: (a) to an ice cream shop, which required \textasciitilde{4} heading changes and \textasciitilde{12} steps (or 1 jump and 3 steps) and virtually traveling 107 meters (350 ft) (b) and to a Japanese restaurant in Paris, which required \textasciitilde{7} heading changes and \textasciitilde{5} steps (or 1 jump and 2 steps) across 73 meters (240 ft). On the left, the map shows the starting location and heading as an orange circle with the path to the destination in blue.}
  \Description{This multi-part figure shows the tasks from Part 3 (the open-world investigations). There are two rows: one row for each task with the top row for the ice cream navigation task and the bottom row for the Parisian Japanese restaurant task. The first image for both shows the travel map where the users had to go. The rest of the images show example StreetViewAI images and status updates along the way.}
  \label{fig:study-method-nav-task} 
\end{figure*}

For the first navigation task, participants were told to \myquote{virtually investigate a walk from your Airbnb rental to a local ice cream shop and also determine what else is nearby (\textit{e.g.,} is there a nice place to eat your ice cream)} and for the second: \myquote{You are visiting a friend in Paris and meeting at a Japanese restaurant. You are investigating the walk from the nearest bus stop to the restaurant.} See \autoref{fig:study-method-nav-task}. For both tasks, we explained \textit{which} direction to start walking but neither how far nor the end addresses. Moreover, we specifically started the participants with a view facing away from the destination. Notably, these tasks were designed as open-world navigation rather than routing tasks because the system did not know the user's destination (we return to routing in the Discussion). 

\subsection{Data and Analysis}
With participant consent, all sessions were audio and video recorded using both room- and screen-based recordings. To monitor tool usage, we built a lightweight, client-side logger that recorded user commands and AI interactions (including prompts, responses, and system actions) as timestamped JSON files. The lead researcher took notes throughout each session, observing navigation behaviors, user reactions, and marking when the AI provided an inaccurate response or refused to answer. These notes were corroborated \textit{post hoc} via log analysis and a review of the video recordings. After each task, participants were asked to reflect on their experience and respond to 7-point Likert-scale questions on ease-of-use, information value, and perceived accuracy (7 is best). Participant responses were analyzed via thematic analysis~\cite{Clarke_ThematicAnalysis_JPP17}.

Because study participants were using an unfamiliar keyboard and OS, we added tactile markers to the hot keys to improve usability. Still, five of the eleven participants inadvertently hit a key that closed \sysname during the study: P8 once and P2, P5, P7, P10 twice each. Because our custom logger batches log data to persistent storage every ten entries, we consequently experienced mild data loss (\textasciitilde{}90 log entries total; < 0.6\% of all logged events).

\section{Study Findings}
Overall, participants reacted positively to the prototype: \myquote{Most navigation systems can get you to the last 5-10 feet but this helps you get to a door and even describes that door} (P10), \myquote{If I'm going to a place, I can familiarize myself first from my home} (P1), and \myquote{This is incredible!} (P5). All eleven participants completed the POI investigations and 10/11 completed at least one open-world navigation task (P9 left before this study task began because his \textit{Access Transit} arrived early). In total, participants moved to 356 panos, made 568 heading changes, and 1,053 AI requests (136 AI Describer and 917 AI Chats). 

During the post-study debrief, participants rated the overall usefulness of \sysname as 6.4/7 (\textit{Median=}7; \textit{SD=}0.9) and all wanted to use the system as a product, if available (\textit{Avg=}6.6; \textit{Med=}7; \textit{SD=}0.8): \revised{\myquote{Is this available? I want to use it now!} (P10) and \myquote{I'm very excited for this to come out, it's going to make a lot of blind people very happy} (P5)}. While participants found value in \sysname, relied heavily on its AI features, and adeptly combined virtual world navigation with AI interactions, they occasionally struggled with orientation, distinguishing the veracity of AI responses, and determining the limits of AI knowledge. Below, we first describe AI usage and performance followed by how participants accomplished the study tasks and high-level themes.

\subsection{AI Usage and \revised{Technical Performance}}
Across the study tasks, participants used AI 1,053 times with a strong preference towards \textit{AI Chat} (917 chats) \textit{vs.} 136 \textit{AI Describe} invocations\footnote{Participants had a total of 996 interactions with AI chat; however, 52 were inadvertent ``live mic'' instances (\textit{e.g.,} where the system overheard a participant's ``think-aloud'' comment or a remark from the study facilitator) and 27 were misheard requests.}. On average, participants used AI Describe 12.4 times (\textit{Median=}10; \textit{SD=}7.5) \textit{vs.} AI Chat 93.0 times (\textit{Med=97.0}; \textit{SD=}37.0) per session. As P10 said, \myquote{The more you ask, the more you learn.} The AI response times were fast, on average less than a second (\textit{Avg=}968.6 ms; \textit{Med=}723; \textit{SD=}678.8) with the AI Chat system---which uses Gemini's Live Multimodal API---responding faster (\textit{Avg=}842 ms; \textit{SD=614.3}) than AI Describe (\textit{Avg=}1,916 ms; \textit{SD=265.6}). Interestingly, a large majority of AI Chat questions were voiced (over 94.4\%) perhaps due to a combination of participant unfamiliarity with the Mac keyboard, the flexibility and ease of voice, and the affordances of multimodal input (\textit{i.e.,} keyboard + AI Chat simultaneously). Below, we analyze question types, AI response accuracy, and non-responses (\textit{e.g.,} \myquote{I'm sorry, I cannot answer that.}). Quotes are from actual AI chat sessions in the study.



\begin{table}[htbp] 
\centering        
\scriptsize       
\setlength{\tabcolsep}{2pt} 
\renewcommand{\arraystretch}{1.1} 

\begin{tabular}{@{} >{\raggedright\arraybackslash}p{2.2cm} >{\raggedleft\arraybackslash}p{0.5cm} >{\raggedleft\arraybackslash}p{0.5cm} >{\raggedright\arraybackslash}p{4.8cm} @{}}
\toprule
\textbf{Question Type} & \textbf{Cnt} & \textbf{\%} & \textbf{Example Questions} \\
\midrule

Spatial orientation & 248 & \cellcolor[RGB]{8,48,107}\color{white}\textbf{27.0\%} & \myquote{How far is the bus stop from where I'm standing?}, \myquote{How close are the garbage cans to the benches?}, \myquote{What's to the left of this building?} \\

Object existence & 243 & \cellcolor[RGB]{8,48,107}\color{white}\textbf{26.5\%} & \myquote{Is there a merry-go-round here?}, \myquote{Are there soccer fields?}, \myquote{Is there a bench at the park?} \\

Description requests & 169 & \cellcolor[RGB]{41,108,162}\color{white}\textbf{18.4\%} & \myquote{What's in front of me?}, \myquote{What view do you see?}, \myquote{What do you see now?} \\

Object/place location & 137 & \cellcolor[RGB]{57,134,189}\color{white}\textbf{14.9\%} & \myquote{Where is the nearest intersection?}, \myquote{Where is the restaurant entrance?}, \myquote{Can you help me find the door?} \\

Followup questions & 132 & \cellcolor[RGB]{57,134,189}\color{white}\textbf{14.4\%} & \myquote{Tell me more about the mural}, \myquote{Is the sidewalk connected to the path?} \\

Nearby places & 85 & \cellcolor[RGB]{97,173,213}\textbf{9.3\%} & \myquote{Is there a train station close by?} \\

Niceties/gratitude & 73 & \cellcolor[RGB]{123,183,218}\textbf{7.9\%} & Expressions of gratitude or conversational niceties \myquote{Thank you.} \myquote{That's interesting!}\\

Directions & 67 & \cellcolor[RGB]{123,183,218}\textbf{7.3\%} & \myquote{How do I get to the door without bumping into any chairs or tables?} \\

Clarifications/confirmations & 59 & \cellcolor[RGB]{149,197,227}\textbf{6.4\%} & \myquote{When you say the crosswalk is directly in front of me, what am I currently facing?}, \myquote{What do you mean by bollards?} \\

Safety/accessibility & 51 & \cellcolor[RGB]{149,197,227}\textbf{5.6\%} & \myquote{Is there an assistive pedestrian signal at this light?}, \myquote{Is there a marked crosswalk at this street?}, \myquote{Are there stairs along this route?} \\

Commands & 40 & \cellcolor[RGB]{174,210,235}\textbf{4.4\%} & Moving, turning, jumping commands (\textit{e.g.,} \myquote{Zoom into the bus stop sign to read the schedule}) \\

Entrance-related & 30 & \cellcolor[RGB]{199,223,243}\textbf{3.3\%} & \myquote{What does the door look like?}, \myquote{How do I get to the entrance?}, \myquote{Are there obstacles to the door?} \\

Additional info requests & 29 & \cellcolor[RGB]{199,223,243}\textbf{3.2\%} & \myquote{Can you tell me what's on the menu?}, \myquote{What bus routes serve this stop?} \\

State/condition inquiries & 19 & \cellcolor[RGB]{224,236,251}\textbf{2.1\%} & \myquote{Does this [street/sidewalk/restaurant] look busy?} \\

Signage reading & 15 & \cellcolor[RGB]{239,246,255}\textbf{1.6\%} & Requests to read signs or text \\

\bottomrule
\end{tabular}
\caption{\revised{Findings from our analysis of AI Chat across all study tasks (N=917). Each user question was annotated with up to three thematic tags drawn from an emergent list of 23 question-type categories.}}
\label{tab:question_types_analysis}
\end{table}

\textbf{Question types.} We analyzed all 917 AI Chat interactions and annotated each with up to three tags drawn from an emergent list of 23 question-type categories. Most commonly, with 27.0\% of interactions (248/917), participants asked about spatial orientation of an object or themselves (\textit{e.g.,} \myquote{How far is the bus stop from where I'm standing?}, \myquote{How close are the garbage cans to the benches?}) followed closely by asking about object existence (243; 26.5\%), including sidewalks, obstacles, and doors---\textit{e.g.,} \myquote{Is there a merry-go-round here?}, \myquote{Are there soccer fields?}, and \myquote{Is there a bench at the park?}. Third were questions asking for a description (169; 18.4\%) such as \myquote{What's in front of me?} or \myquote{What view do you see?} followed by where an object or place is (137; 14.9\%)---\textit{e.g.,} \myquote{Where is the nearest intersection?}, \myquote{Where is the entrance to the restaurant?}, or \myquote{Can you help me find the door?}. Demonstrating the conversational nature of AI Chat, over 14\% (132) of chat interactions were followups such as \myquote{Tell me more about the mural} or \myquote{Is the sidewalk connected to the path?} or non-questions expressing gratitude or niceties (7.9\%). Asking for specific directions (67; 7.3\%) and about nearby places (85; 9.3\%) was also common:\myquote{How to get to the door without bumping into any chairs or tables?} and \myquote{Is there a train station close by?}

Interestingly, over 6\% of chats (59) were clarifications or confirmations (\textit{e.g.,} \myquote{When you say the crosswalk is directly in front of me, what am I currently facing?}, \myquote{What do you mean by bollards?}, or \myquote{Just to confirm, there is no bike rack along the sidewalk?}) as participants sought to strengthen their understanding or double check previous information. Many of the existence and location inquiries were related to safety or accessibility (51; 5.6\%) (\textit{e.g.,} \myquote{Is there an assistive pedestrian signal at this light?}, \myquote{Is there a marked crosswalk at this street?}, \myquote{Are there stairs along this route?}), 30 (3.3\%) were related to entrances (\textit{e.g.,} \myquote{What does the door look like?}, \myquote{How do I get to the entrance?}), and 15 (1.6\%) were requests to read signage. 

Other unexpected but less common questions emerged such as asking about the state of something (19; 2.1\%), such as \myquote{Does this [street/sidewalk/restaurant] look busy?}), asking questions that would require additional information sources beyond what we currently supply the model (29; 3.2\%), such as \myquote{Can you tell me what's on the menu?} or \myquote{What bus routes serve this stop?}. Over 4\% (40) of chat interactions were commands, including moving, turning, and jumping but not all of which the AI could perform. \revised{For example, P10 said: \myquote{Zoom in to read the bus schedule} in reaction to the AI explaining it could not read the sign because it was too far away; however, ``zoom'' is not currently implemented in \sysname}.

\textbf{Accuracy}. Of the 816 questions asked, 703 (86.3\%) were correctly answered, 32 were incorrect (3.9\%), 26 were partially correct (3.2\%) and, for the the remaining 54 (6.6\%), the AI told the user it could not provide an answer. Of the 32 incorrect responses, 20 (62.5\%) were false negatives---\textit{e.g.,} stating that a bike rack or park bench did not exist when it did, four (12.9\%) were misidentifications (\textit{e.g.,} a yellow speed bump interpreted as a crosswalk, a distant sign as a gate), and another four (12.9\%) were errors that occurred when the user asked about the existence of something before the destination was fully in view (\textit{e.g.,} asking if outdoor seating existed before being close enough to the restaurant to clearly see it). Though uncommon, two other interesting errors emerged: correctly identifying an object or place but misrepresenting its location (\textit{e.g.,} the AI told one participant that the Japanese restaurant was on the corner but it is located on the middle of a block) and being told that you could move closer to a destination but you could not. Here, the AI model confused being physically present in a scene---where you could rightly walk forward to the destination but in \sysname, due to the availability of GSV imagery, the participant was as close as possible. This understandably confused the participant.

\textbf{Refusals.} For the 54 queries that the AI refused to answer, it was most often because a request required geographic knowledge the AI did not have access to (25 of 54; 46.3\%) such as asking about the nearest transit stop or getting directions to a place not in the immediate search radius (50 meters). Note that we did not connect the AI model to the full Google Places API (or other geographic knowledge sources), so its geographic data was limited. Similarly and second most commonly, participants asked questions that required additional knowledge bases such as available bus routes, transit schedules, or restaurant menus at particular locations (15/54; 27.8\%). The remaining non-responses were due to unreadable signage, objects the AI couldn't see, or questions beyond the AI's capabilities. For example, one participant asked about relative distances between playground equipment, which resulted in the response \myquote{I'm sorry, I cannot tell you exact distances between playground features}, while another asked the AI agent to actually plan his visit to the playground with his nieces (\myquote{I cannot plan your specific visit...}).

We now turn towards describing the POI and open-world navigation tasks before enumerating some high-level themes.

\subsection{POI Investigation}
For each POI investigation, participants were told the task-specific scenario, teleported directly to the destination, and asked to explore on their own given the task (\textit{e.g.,} \myquote{investigate the playground for safety and fun}). At the end of each task, participants were asked questions about the location and told that if they did not immediately know the answer, they could use \sysname to find out. To avoid influencing reactions to our prototype, the study facilitator did not reveal answer correctness until the debrief.

To accomplish the tasks, participants employed a variety of techniques but primarily relied on AI for information. Upon arrival, they listened carefully to the teleport message, which provided location data, nearby places, the user's heading, and possible movements. Most then began with an AI Describe description (\keys{Alt} + \keys{D}) before utilizing AI Chat (\keys{Alt} + \keys{C}) to drill down. Others simply started with chat. Overall, participants used AI Describe an average of 7.3 times per participant (\textit{Med=}8.0; \textit{SD=}3.4) and AI Chat 57.3 times (\textit{Med=}55.0; \textit{SD=}23.0) across the four locations (including the tutorial). While we instructed participants that it was not necessary to move or pan to complete the POI tasks (except for the playground task), most at least changed their heading at each location (\textit{Avg=7.8}; \textit{Med=}5), demonstrating an interest in looking around to gain more scene understanding. As P8 said:

\begin{myblockquote}
This is literally what I do when I physically relocate to a new place or work at a new building, I try to build a whole mental model of everything that is around me so that I can orient myself. So, I think this would be very useful for that use case and, also, obviously for planning to go to unfamiliar places.
\end{myblockquote}


\textbf{Post-task questions.} Impressively, all participants successfully answered the post-task questions with a high correctness rate (\autoref{tab:poi_participant_answers}). Of the 20 scene-related questions, participants correctly answered 18.6 correctly (93\%). Across all 220 questions asked (20 x 11 participants), there were 9 total incorrect answers, 4 partially correct, and 3 non-responses. All mistakes were due to inaccuracies in the AI and not due, for example, to participant misunderstanding. The most common AI error was stating that there was \textit{not} a bike rack at the Mexican restaurant (there was, though it was not particularly salient), that there \textit{was} outdoor seating at the restaurant (there wasn't), or that there were \textit{not} benches at the park (there was but they were in the distance). Partial errors involved reporting only grass at the playground (there was also wood chips). Interestingly, as demonstrated by \autoref{tab:poi_participant_answers}, the AI was reliable but randomly inconsistent across participants---responding accurately most of the time but in a small number of cases, failing (where it had previous succeeded) or omitting details. We return to accuracy perceptions in the themes below and in the Discussion.

\begin{table}[htbp] 
\centering        
\footnotesize     
\setlength{\tabcolsep}{3pt} 
\renewcommand{\arraystretch}{1} 
\begin{tabular}{@{} >{\raggedright\arraybackslash}p{3.5cm} >{\raggedright\arraybackslash}p{3.7cm} >{\centering\arraybackslash\renewcommand{\arraystretch}{1.0}}m{0.8cm} @{}}
\toprule
\textbf{Questions} & \textbf{Correct Response} & \textbf{Correct} \\
\midrule
\multicolumn{3}{@{}l}{\textbf{Bus Stop}} \\
What does the bus stop look like? & \myquote{A covered bus stop with benches on a university campus} & \cellcolor[RGB]{8,48,107}\color{white}100\% \\
Are there benches? & \myquote{Yes} & \cellcolor[RGB]{8,48,107}\color{white}100\% \\
Is there a bus shelter? & \myquote{Yes} & \cellcolor[RGB]{8,48,107}\color{white}100\% \\
What are some surrounding navigational landmarks? & \myquote{Library, student union, bus stop shelter, and garbage cans to the left} & \cellcolor[RGB]{8,48,107}\color{white}100\% \\
Are there garbage cans? & \myquote{Yes} & \cellcolor[RGB]{8,48,107}\color{white}100\% \\
What are some nearby buildings? & \myquote{Student union, library, \textit{etc.}} & \cellcolor[RGB]{8,48,107}\color{white}100\% \\
\midrule
\multicolumn{3}{@{}l}{\textbf{Playground}} \\
How does the playground look? & \myquote{A playground with swings, slides, and a climbing structure} & \cellcolor[RGB]{8,48,107}\color{white}100\% \\
Are there slides? & \myquote{Yes} & \cellcolor[RGB]{8,48,107}\color{white}100\% \\
Are there swings? & \myquote{Yes} & \cellcolor[RGB]{8,48,107}\color{white}100\% \\
What kind of ground cover? & \myquote{Grass with wood chips under the play structure} & \cellcolor[RGB]{76,149,198}77.3\% \\
Are there benches to sit on? & \myquote{Yes} & \cellcolor[RGB]{57,134,189}\color{white}81.8\% \\
Are there sidewalks? & \myquote{Yes} & \cellcolor[RGB]{8,48,107}\color{white}100\% \\
Does the street look busy? & \myquote{No} & \cellcolor[RGB]{8,48,107}\color{white}100\% \\
What kind of neighborhood? & \myquote{Residential} & \cellcolor[RGB]{8,48,107}\color{white}100\% \\
What kind of houses are nearby? & \myquote{Two-story, single-family homes} & \cellcolor[RGB]{8,48,107}\color{white}100\% \\
\midrule
\multicolumn{3}{@{}l}{\textbf{Mexican Restaurant}} \\
What does the building look like? & \myquote{2-story brick building w/a blue awning, yellow side, \& a mural} & \cellcolor[RGB]{76,149,198}77.3\% \\
What's unique re: the building? & \myquote{A mural. Large, green creature (dragon) with purple hair.} & \cellcolor[RGB]{8,48,107}\color{white}100\% \\
Is there a parking lot nearby? & \myquote{Yes} & \cellcolor[RGB]{30,93,152}\color{white}90.9\% \\
Is there a bike rack? & \myquote{Yes} & \cellcolor[RGB]{123,183,218}63.6\% \\
Is there a sidewalk out front? & \myquote{Yes} & \cellcolor[RGB]{8,48,107}\color{white}100\% \\
\bottomrule
\end{tabular}
\caption{Post-task scene related questions and the correctness of participant answers. To address these questions, participants primarily relied on AI Chat but also AI Describe and the hotkey \keys{Alt} + \keys{N} for nearby places.}
\label{tab:poi_participant_answers}
\end{table}


\subsection{Open-world Navigation}
For the open-world navigation tasks, participants were first asked to find an ice cream shop in Minneapolis and then a Japanese restaurant in Paris (\autoref{fig:study-method-nav-task}). To accomplish the ice cream task, the participant needed to shift their view to the right (east) 90° and then walk (or jump) 350 feet to the first intersection. At the intersection, Ben and Jerry's was on the participant's left (north). For the Japanese restaurant task, the user needed to turn around (180°) to face north then walk or jump 130 feet (40m) to the nearest intersection and turn left (west), then walk 75 feet (23m) with the destination on the right. Once the user thought they had arrived at their destination, they were asked to answer questions relevant to the task (\textit{e.g.,} \myquote{Is there a nice place to walk with their ice cream?} or \myquote{What food was advertised at the Japanese restaurant?}).

\textbf{Completing the tasks.} All ten participants who began the ice cream task finished it but with varying strategies and efficiency. On average, it took participants 13.8 minutes with 10.2 movements, 28.3 heading changes, and 17.2 AI invocations (3.5 AI Describe; 13.7 AI Chat). The most successful participants listened intently to each status message (which occur at panning/moving), turned and virtually walked towards the ice cream shop until they heard ``Ben and Jerry's'' as a nearby place, and then continued stepping forward while asking the AI about the shop's specific location and whether it was visible. Those who struggled either walked or jumped past the ice cream shop and got confused or overly relied on nearby place descriptions (\keys{Alt} + \keys{N}) at a cost of involving AI Chat. Due to limited time, only 7 of the 11 participants conducted the Japanese task; all finished. Here, it took participants an average of 9.4 minutes with fewer movements and AI uses than the ice cream shop, perhaps because of the shorter distance or increased experience of the participants.



\textbf{Own locations.} Six participants had time to select and explore their own location; most chose their own homes or a favorite restaurant. One person selected a place where they studied abroad 25 years ago in England, another selected a home he grew up in, and a third explored the walk from their apartment to the \textit{Department of Services of the Blind}. All participants enjoyed listening to the AI-generated descriptions, comparing them to their own understandings: P6 said, \myquote{See, that's my house. And I could turn right here to get to 168th St.} (which she then did) and P1 noticed a difference, \myquote{Oh, this image must have been taken when we still had a garage.} P11 visited \textit{Chick-fil-A}. After teleporting and hearing the description, he instantly recognized that \myquote{It took us to the wrong one} as there were two Chick-fil-A's on the same street but a few miles apart. Similarly, P5 was able to discover that \sysname brought him to the back of his father's apartment building rather than the front and then used \sysname to virtually walk around the building.  

\subsection{Overarching Themes}
At the end of the study, participants were asked debrief questions, including \textit{``How well did \sysname support you in finding the information you needed?''}, \textit{``How valuable was the information \sysname provided?''}, and \textit{``How accurate did you perceive the information?''}---see \autoref{tab:likert-results-overview}. In general, participants felt that the tool supported them in finding information and that the information itself was valuable but had some concerns about accuracy. Below, we report on cross-cutting themes extending the sections above.

%
%
%
%
%

\begin{table}[htbp]
\centering
\footnotesize
\renewcommand{\arraystretch}{1.1}
\setlength{\tabcolsep}{3pt}
\begin{tabular}{@{}lcccccc@{}}
\textbf{} & \multicolumn{3}{c}{\textbf{POI Investigation}} & \multicolumn{3}{c}{\textbf{Navigation}} \\
\cmidrule(lr){2-4} \cmidrule(l){5-7}
 & Mean & Median & SD & Mean & Median & SD \\
\midrule
\rowcolor[HTML]{EFF2F6}
How well & \cellcolor[RGB]{44,123,182}\color{white}6.0 & \cellcolor[RGB]{44,123,182}\color{white}6.0 & 0.7 & \cellcolor[RGB]{44,123,182}\color{white}6.0 & \cellcolor[RGB]{44,123,182}\color{white}6.0 & 0.9 \\
\rowcolor[HTML]{FFFFFF}
How valuable & \cellcolor[RGB]{8,81,156}\color{white}6.4 & \cellcolor[RGB]{8,48,107}\color{white}7.0 & 0.6 & \cellcolor[RGB]{44,123,182}\color{white}6.1 & \cellcolor[RGB]{8,81,156}\color{white}6.5 & 1.1 \\
\rowcolor[HTML]{EFF2F6}
How accurate & \cellcolor[RGB]{107,174,214}5.7 & \cellcolor[RGB]{44,123,182}\color{white}6.0 & 0.9 & \cellcolor[RGB]{158,202,225}5.5 & \cellcolor[RGB]{107,174,214}5.5 & 1.2 \\
\bottomrule
\end{tabular}
\caption{\textit{Post-hoc} 7-point Likert scale ratings (7 is best).}
\label{tab:likert-results-overview}
\end{table}

\textbf{Perceptions of accuracy}. Interestingly, participants generally rated \sysname's accuracy as high (\textit{Med=}6.0 for POI investigation and 5.5 for navigation) even though, as P1 said, \myquote{It seems pretty accurate but I don't have vision so I don't really know} and P10 said \myquote{It didn’t come up with hallucinatory answers---as far as I know}. These perceptions were likely influenced by the intermixing of two data sources, geographic data from Google Maps (which is highly accurate) and AI descriptions (which may not be). For example, P11 experienced a problem in the POI task where the AI chat said that there was no Mexican restaurant nearby, however, as they rotated their view, they heard the restaurant reported in the list of local nearby places close to them (which is not inferenced). \myquote{She [the AI] got confused about the Mexican restaurant. But when I was spinning around, she did a great job describing what was around.} Similarly, P8 felt that the tool was hallucinating but then used hotkeys to confirm information about the current road and nearby places.

\textbf{Conceptualizing street view and controls.} As we saw with our co-design, it is challenging to explain the concept of high-resolution, immersive 360° streetscape imagery and navigating therein to blind users. After the third participant, we changed our study introduction to incorporate role play: participants physically stood up and rotated their bodies 45° to the right (to demonstrate \keys{{$\rightarrow$}}) and 45° to the left (for \keys{{$\leftarrow$}}). Participants also took a physical step along their current heading and compared this to virtual steps in \sysname (which were 10-15 meters). This strengthened understanding and led to fewer confusions, especially for initial tasks. We return to disembodied control in the Discussion.


\textbf{Building spatial awareness.} As evident by the number of orientation-related AI chats, participants focused heavily on building a mental model of a space---where the person was inside the virtual-physical world, where objects were, and there relative positions. \revised{Though imperfect, the MLLM surprised us in some cases; for example, in imagining new orientations and providing guidance}:

\begin{myblockquote}
\textbf{P9:} Let's say I stepped out of the restaurant, can you orient me from that perspective?

\textbf{AI:} Okay, if you were to step out of the restaurant, you would be facing the street. You would have the Thompson St. sign on your left and be facing Southeast.
\end{myblockquote}

The interactive chat allowed participants to orient their perspectives and strengthen their mental models:

\begin{myblockquote}
\textbf{P10:} Where is the door?

\textbf{AI:} The door to the cafe is directly in front of you.

\textbf{P10:} Do I have to cross the street to get to it?

\textbf{AI:} No, you do not have to cross the street to get to the door. It is on your side of the street.

\textbf{P10:} Is it on my left or right side?

\textbf{AI:} The door is directly in front of you, not to your left or right.

\textbf{P10:} That's interesting! I thought the intersection was in front of me.

\textbf{AI:} You're right, the intersection is also straight ahead, and the door is just to the left of that. So, while the intersection is straight ahead, the door is slightly to your left relative to the intersection.
\textbf{}

\end{myblockquote}

Still, because \sysname intermixes both relative (\myquote{On your left...}) and absolute positions (\myquote{You're facing northeast}), some participants struggled to make sense of this information, especially since panning status messages always used absolute references.

\revised{\textbf{Alternative access to information.} Finally, we asked participants \myquote{If you didn't have a tool like \sysname, how would you otherwise find out the information it provides}. P3 said, starkly, \myquote{I wouldn't.} but most participants mentioned relying on others: \myquote{I would bring someone with me [to the location] and ask them all the questions} (P4), \myquote{Using paratransit and asking people} (P6), or \myquote{Asking people on AIRA~\cite{aira_video_remote_interpreting}---[a remote video interpreting service]}. P7, P8, and P10 mentioned existing accessible \textit{in situ} navigation tools like \textit{Audio Guidance} in Google Maps \cite{Google_GoogleMapsVoiceGuidance_2023} or VoiceVista~\cite{VoiceVista} and walking around physically. In comparison to those tools, however, P10 emphasized: \myquote{What I like about this, is that I can ask questions and be proactive---that's a real game changer for me.}}


\section{Discussion}
We introduced and evaluated \textit{\sysname} (\autoref{fig:teaser}), the first accessible street view tool for BLV users using context-aware AI and keyboard-based navigation controls. Our findings help highlight \textit{what} information BLV users desire from and ask about streetscape imagery, the potential of multimodal AI models to appropriately answer contextualized inquiries, and uncovers important implications for the design of future accessible streetscape tools. We reflect on key findings, limitations, and opportunities for future work.

\subsection{Interacting with AI-Powered Streetscapes}
Our study revealed valuable insights into how blind users engage with and perceive an AI-driven streetscape environment. A key finding was the participants' preference for the AI Chat Agent over passively receiving triggered descriptions (AI Describer via \keys{Alt} + \keys{D}), suggesting a desire for agency and targeted information seeking. Unlike AI Describer, the chat system is intrinsically personalized to questions of interest. However, the reliance on AI for sensemaking and navigating complex visual scenes also surfaced potential issues around accuracy and trust, data sources and discrepancies, and the ability for the AI backend to go beyond the user's current context to address questions.

\textbf{Accuracy and trust}. Participants exhibited a high-level of trust in \sysname, even with inaccuracies. As other scholars have emphasized~\cite{Kaate_YouAlwaysGetAnAnswer_IUI25, Yin_DoLLMsKnowWhatTheyDontKnow_ArXiv2023, Nahar_FakesOfVaryingShades_ArXiv2024}, it is difficult to discern hallucinations from true information---\sysname seems just as confident about both. For example, \sysname repeatedly told participants that there was no bike rack at the Mexican restaurant but there was (though in some sessions, it got it right). \sysname's ability to respond and provide human-like answers for most questions---whether they were correct---increased perceptions of trust and accuracy: \myquote{There are so few things that the system seemed like it could not answer} (P10). \revised{While the risks in a virtual environment are lower here than relying on real-time AI guidance in the physical world, the potentially flawed mental models of geography and spatial relationships could still have profound effects on planning travel and safety.} Future work should explore how best to frame information gained from \sysname and limitations therein, borrowing from literature in Explainable AI (XAI)~\cite{Arrieta_ExplainableAI_InfoFusion2020}.


\textbf{Discrepancies in data.} \sysname draws primarily on two information sources: geographic databases of road, place, and address information (which is up-to-date and trustworthy) and AI-generated descriptions about the scene and local geography (which can depend on outdated street view imagery and imperfect inferences). Occasionally, these data sources would disagree: \textit{e.g.,}, for one participant, the AI said that there was \myquote{no Mexican restaurant in front of them} even though the geographic-based status message said that there was. Interestingly, when specifically asked about whether participants \textit{thought} about streetscape image age when using \sysname in the post-study debrief, only a few did---despite being taught how to use \keys{Alt} + \keys{P} to hear metadata about the GSV photograph, including capture date.

\textbf{Increasing data sources.} Over 70\% of non-responses from the AI (40/54) were due to a lack of connected data sources, including transit schedules, nearby places beyond the immediate area (> 50 meter context radius), and restaurant menus. These queries should all be possible in the future by feeding additional data into the prompts or by connecting the multimodal AI to external knowledge sources (\textit{e.g.,} via function calls). Similarly, to address limitations in streetscape image age, future work should incorporate additional image datasets (\textit{e.g.,} the Google Places API contains millions of business- and user-contributed photos~\cite{google_maps_places_photos_api}, which are often up-to-date and show both indoor and outdoor imagery).

\textbf{Towards a more autonomous AI agent.} While the AI Chat Agent had access to all previous user views---from heading changes and movement---in its context window (up to \textasciitilde{1M} input tokens), \sysname could not address queries that went beyond these views unless that data was provided as geographic context (\textit{e.g.,} nearby places within 50 meters, nearby roads, \textit{etc.}). Future work should create a more autonomous agentic backend capable of examining the full 360° at every location and all nearby panos and metadata to address user queries. For example, \myquote{What's the next bus stop down this road and what does it look like?}. Here, the AI agent should automatically query geographic knowledge sources, find the bus stop location, capture and analyze surrounding GSV imagery, and return a response. The user could then inquire about the walking route to the bus stop, which the AI agent could again automatically analyze using both geographic metadata as well as CV analysis of GSV images along the way. 

\subsection{Navigation, Orientation, and Routing}\label{sec:disc_orientation}
Effectively navigating the virtual environment itself presented distinct challenges and opportunities for our blind participants, including user orientation, precise placement within the scene, and desired pathfinding capabilities.

\textbf{Orientation}. In a virtual, disembodied environment, supporting and maintaining the user's understanding of their orientation with respect to the current location, view, and spatial relationships therein is a significant challenge. We found that standing and physically role playing turns \textit{vs.} steps helped develop appropriate mental models of the movement controls but that maintaining spatial orientation was difficult. Some users strongly preferred relative directions only (\textit{e.g.,} \myquote{On my left...} or \myquote{In front of me...} despite absolute directions (cardinal and intercardinal headings) having clear advantages of being non-contextual. These challenges may fade as users gain more experience.

\textbf{The importance of landing.} When teleporting to an address or POI, we used a custom algorithm to automatically turn the user's view towards that destination; however, this technique was imperfect and the cost of being wrong and recovery time for a blind user \textit{vs.} a sighted user is disproportionately high. Even if the heading angle was off by a small amount (\textasciitilde{45°}) or a sub-optimal pano was selected (\textit{e.g.,} 10 meters from the destination), the blind user needs to first \textit{discover} the error (\textit{e.g.,} \myquote{I'm not looking at what I thought...}) and then recover (\textit{e.g.,} \myquote{Which direction should I be facing? Where is the destination from here?}). Future work should use CV analysis of the surrounding drop points and determine the best pano and view for a blind user to investigate the destination (\textit{e.g.,} a store's entrance and the pedestrian pathway in front), or could ask the user which of several options they prefer.

\textbf{Routing.} Building on the above, while \sysname currently supports POI investigations, open-world navigation, and virtual tourism, it does not yet support origin-to-destination routing---a key task identified in our co-design sessions. In contrast to open-world navigation where \sysname knows the origin but not the destination, in routing, the tool would know both. At a minimum, supporting this feature would enable users to virtually travel a route with turn-by-turn navigation (similar to \textit{Voice Guidance} in Google Maps~\cite{Google_GoogleMapsVoiceGuidance_2023}). As P9 said, \myquote{I want to ask about the 'last mile' walk from the nearest bus stop to the cafe}. Other interesting features are also worth exploring, such as: (1) using an AI agent to examine all panos along potential routes and determining potential obstacles; (2) creating blind-friendly summaries of the route; and (3) helping with the ``last 10 meters''~\cite{Saha_LastFewMetersWayfindingsForPeopleWithVI_ASSETS2019}, including door finding.

\textbf{Bookmarking.} Finally, similar to \textit{in situ} navigation tools such as \textit{VoiceVista}~\cite{VoiceVista} and \textit{FootNotes}~\cite{Gleason_FootNotesGeoReferencedAudioAnnotations_IMWUT18}, users wanted the ability to bookmark locations, annotate them, and then easily retrieve them either again on the laptop or while actually physically navigating in the real world on their smartphone.

\subsection{Audio Descriptions and Beyond}
In {\sysname}, the primary output is discrete verbalized messages. We reflect on content, concision, and other possibilities.

\textbf{Ephemeral data.} In the post-study debrief, participants emphasized that certain vocalized details were irrelevant and distracting, such as ephemeral objects (\myquote{That parked car would not still be there though, right?}), car colors, or the presence of pedestrians or traffic. While sighted users of GSV can easily ignore these visual details---and, indeed, may not even notice them---they take up valuable audio bandwidth if described by the AI model. Still, some participants felt that, with caveats, such information was useful to determine, for example, how busy a location could be. \revised{Future tools should enable users to customize what features are most important to describe and what to ignore.}

\textbf{Concise announcements.} To maintain user orientation and support their mental model of the local geography and navigable areas, we provide status updates at every interaction from heading changes to movement; however, constructing these messages for clarity, relevance, and concision is challenging. In initial prototypes, we triggered AI Describer descriptions at every heading change or movement. In later prototypes, including that used in our study, we decided to require explicit hot key presses (\keys{Alt} + \keys{D}). Still, these status updates could be too verbose, which would cause frustration and lead to information loss (as important messages are diluted). Participants suggested being able to customize the output for their current tasks both in terms of what the AI describes as well as what is included in the status updates.

\textbf{Going beyond audio descriptions.} As an initial prototype, we relied on discrete audio descriptions for both navigational feedback and AI descriptions (earcons were used to notify the user about microphone toggling). However, many accessible navigation tools~\cite{Microsoft_Soundscape_2023, BlindSquare, Walker_NavPerfWithVirtualAuditoryDisplay_HumanFactors2006}, VR techniques~\cite{Schloerb_BlindAid_IEEEHaptics10,Wedoff_VirtualShowdownAccessibleVR_CHI19, May_SpotlightsAndSoundscape_TOCHI20, Siu_VRWithoutVision_CHI20}, and audio games~\cite{AFB_ShadesOfDoom, Atkinson_AudioQuake_GitHub18} use spatialized audio or haptics to provide non-verbal cues about distances, object locations, and other information. Future work should examine these additional, complementary output modalities (\textit{e.g.,} modulating a musical pitch for heading orientation). Unlike digital content in VR or gaming---where exact distances can be trivially calculated---it is more challenging in streetscape images; however, new monocular depth estimation algorithms from RGB images ~\cite{Wang_SingleImageTo3D_ECCV24, Yang_DepthAnything_NEURIPS24} could be used to auto-generate 3D audio soundscapes. Moreover, the sounds of the streetscape could be synthesized and spatially rendered based on emerging image-to-audio synthesizes~\cite{Su_SonifyAR_UIST24, Zhou_VisualToSound_CVPR18} (\textit{e.g.,} hearing the sounds of a cafe or street traffic when turning towards it). 


\textbf{Geo-spatial voice agents for all.} While our primary goal was to create an accessible streetscape tool, as development progressed and we added bidirectional voice conversation capabilities and voice commands, we realized that \sysname could be the beginning of a generalizable audio-only streetscape experience. Imagine, for example, conversing with a voice agent like \textit{Alexa} or \textit{Siri} while driving and learning about the color of a building at an upcoming turn or whether there appears to be a nearby parking lot and where. Such queries could then trigger an AI-based streetscape analysis along the user's route or destination.

\subsection{Limitations}
Our study had three primary limitations. First, while we recruited eleven blind participants, none were familiar with Mac keyboards and some preferred smartphones as their primary computing device; thus, the dominant use of voice \textit{vs.} typing for AI Chat (> 90\%) may be artificially inflated. As P10 said: ``\textit{If I were more comfortable with your keyboard, I would use \keys{Alt}+\keys{D} to get a Gestalt first and then \keys{Alt}+\keys{C} to drill down more}.'' Moreover, while our participants varied in age and demographics, all were cane users and most were technology savvy with medium-to-high AI familiarity. Second, while our question type analysis reveals important findings about \textit{what} information participants are interested in deriving from streetscape imagery---including spatial orientation, the existence or presence of objects, describing the current view---the frequency and nature of questions asked are a reflection of the study tasks as well as our post-task questions. Derived from literature and our co-design sessions, we believe the study tasks had high ecological validity; however, future work should explore real usage of \sysname in a deployment study. Third, and finally, due to study length constraints, we could not examine all parts of the \sysname prototype, including the \textit{AI Tour Guide} and setting a custom user profile. However, these features were co-designed and tested by our blind collaborators; future work should study AI personas and user profiles in more detail.

\section{Conclusion}
\sysname marks a fundamental advance in making immersive streetscape environments accessible. As P8 said, \myquote{Google's Street View tool is not accessible at all... this is a huge leap forward in navigation}. Our work contributes not only a novel system but also design insights for interacting with 360° imagery non-visually, interaction techniques for conversational AI agents in spatial contexts, and a deeper understanding of the information needs of BLV users exploring virtual representations of the physical world.  

\begin{acks}
Diagram icons from Noun Project, including: \href{https://thenounproject.com/icon/ai-prompt-7906362/}{``prompt icon'' by Firdaus Faiz}, \href{https://thenounproject.com/icon/html-7643378/}{``command functions'' by Kawalan Icon}, \href{https://thenounproject.com/icon/place-7892484/}{``<lat,lng>'' by Didik Darmanto}, \href{https://thenounproject.com/icon/compass-4077452/}{``heading'' and ``relative heading'' by IronCV}, \href{https://thenounproject.com/icon/road-7337078/}{``road address'' and ``place address'' by IGraphics}, \href{https://www.flaticon.com/free-icon/earth_18022223}{``city, state, country'' by Karyative}, \href{https://thenounproject.com/icon/cafe-7865395/}{``café'' by loviana}, \href{https://thenounproject.com/icon/hospital-7865371/}{``hospital'' by loviana}, \href{https://thenounproject.com/icon/park-6162285/}{``park'' by Made x Made}, \href{https://thenounproject.com/icon/school-6333109/}{``school'' by Omah Icon}, \href{https://thenounproject.com/icon/storefront-7864275/}{``storefront'' by Sutiyah}, \href{https://thenounproject.com/icon/tags-7862941/}{``place type'' by BEARicons}, \href{https://thenounproject.com/icon/writing-7930041/}{``editorial summary'' by ilham firmansyah}, \href{https://thenounproject.com/icon/ruler-7893407/}{``relative distance'' by Sandy Walsh}, \href{https://thenounproject.com/icon/blind-7764040/}{``blind person'' by Nuno Sequeira}, \href{https://thenounproject.com/icon/joystick-7747276/}{``game controller'' by PMSO BEMFEBUNUD}, \href{https://thenounproject.com/icon/up-key-6141676/}{``up arrow key'' by Se-hui Jo}, \href{https://thenounproject.com/icon/left-key-6141674/}{``left arrow key'' by Se-hui Jo}, \href{https://thenounproject.com/icon/ai-7865331/}{``MLLM icon'' by Funtasticon}.
\end{acks}

\bibliographystyle{ACM-Reference-Format}
\bibliography{refs}


\end{document}